\documentclass[12pt]{article}

\usepackage{latexsym,amsmath,amssymb,amscd,eucal}
\usepackage{xy}
\usepackage{enumerate}

\input xy
\xyoption{all}

\def\harr#1#2{\smash{\mathop{\hbox to .3in{\rightarrowfill}}
 \limits^{\scriptstyle#1}_{\scriptstyle#2}}}

\def\appendix#1{\addtocounter{section}{1}\setcounter{equation}{0}
\renewcommand{\thesection}{\Alph{section}}
\section*{Appendix \thesection\protect\indent \parbox[t]{11.715cm} {#1}}
\addcontentsline{toc}{section}{Appendix \thesection\ \ \ #1} }
\newcommand{\eq}{\begin{equation}}
\newcommand{\eqend}{\end{equation}}

\newbox\ncintdbox \newbox\ncinttbox
\setbox0=\hbox{$-$} \setbox2=\hbox{$\displaystyle\int$}
\setbox\ncintdbox=\hbox{\rlap{\hbox to
\wd2{\hskip-.125em\box2\relax \hfil}}\box0\kern.1em}
\setbox0=\hbox{$\vcenter{\hrule width 4pt}$}
\setbox2=\hbox{$\textstyle\int$}
\setbox\ncinttbox=\hbox{\rlap{\hbox to
\wd2{\hskip-.175em\box2\relax \hfil}}\box0\kern.1em}

\newcommand{\complex}{{\mathbb C}} 
\newcommand{\real}{{\mathbb R}} 

\def\Z{{\mathbb{Z}}}

\def\CS{\mathfrak{S}}

\newcommand{\Tr}[1]{\:{\rm Tr}\,#1}

\def\be{\begin{equation}}
\def\ee{\end{equation}}
\def\bea{\begin{eqnarray}}
\def\eea{\end{eqnarray}}
\def\bd{\begin{displaymath}}
\def\ed{\end{displaymath}}

\DeclareFontFamily{U}{rsf}{}
\DeclareFontShape{U}{rsf}{m}{n}{
  <5> <6> rsfs5 <7> <8> <9> rsfs7 <10-> rsfs10}{}
\DeclareMathAlphabet\Scr{U}{rsf}{m}{n}

\def\cR{{\Scr R}}

\def\cN{{\Scr N}}
\def\cA{{\Scr A}}
\def\cB{{\Scr B}}
\def\cC{{\Scr C}}

\def\cD{{\Scr D}}

\def\cH{{\Scr H}}
\def\cP{{\Scr P}}

\def\cE{{\Scr E}}
\def\cF{{\Scr F}}

\def\cO{{\Scr O}}

\makeatletter
\newdimen\normalarrayskip              
\newdimen\minarrayskip                 
\normalarrayskip\baselineskip
\minarrayskip\jot
\newif\ifold             \oldtrue            
\def\arraymode{\ifold\relax\else\displaystyle\fi} 
\def\@arrayskip{\ifold\baselineskip\z@\lineskip\z@
     \else
     \baselineskip\minarrayskip\lineskip2\minarrayskip\fi}
\def\@arrayclassz{\ifcase \@lastchclass \@acolampacol \or
\@ampacol \or \or \or \@addamp \or
   \@acolampacol \or \@firstampfalse \@acol \fi
\edef\@preamble{\@preamble
  \ifcase \@chnum
     \hfil$\relax\arraymode\@sharp$\hfil
     \or $\relax\arraymode\@sharp$\hfil
     \or \hfil$\relax\arraymode\@sharp$\fi}}
\def\@array[#1]#2{\setbox\@arstrutbox=\hbox{\vrule
     height\arraystretch \ht\strutbox
     depth\arraystretch \dp\strutbox
     width\z@}\@mkpream{#2}\edef\@preamble{\halign \noexpand\@halignto
\bgroup \tabskip\z@ \@arstrut \@preamble \tabskip\z@ \cr}%
\let\@startpbox\@@startpbox \let\@endpbox\@@endpbox
  \if #1t\vtop \else \if#1b\vbox \else \vcenter \fi\fi
  \bgroup \let\par\relax
  \let\@sharp##\let\protect\relax
  \@arrayskip\@preamble}
\makeatother

\newcommand{\beq}{\begin{eqnarray}}
\newcommand{\eeq}{\end{eqnarray}}

\newcommand{\G}{\Gamma}

\def\appendix#1{\addtocounter{section}{1}\setcounter{equation}{0}
\renewcommand{\thesection}{\Alph{section}}
\section*{Appendix \thesection. #1}
\addcontentsline{toc}{section}{Appendix \thesection\ \ \ #1} }

\newtheorem{remark}{Remark}[section]

\newtheorem{comment}{Comment}[section]

\numberwithin{equation}{section}

\begin{document}

\rightline{SISSA 22/2011/EP}

\begin{center}

{\Large\bf Partition Functions for Quantum Gravity, Black Holes, Elliptic Genera and Lie Algebra Homologies}

\end{center}
\vspace{0.1in}

\begin{center}
{\large
L. Bonora $^{(a)}$\footnote{ bonora@sissa.it}\,
and\, A. A. Bytsenko $^{(b)}$
\footnote{abyts@uel.br}


\vspace{10mm}
$^{(a)}$ {\it
International School for Advanced Studies (SISSA),
Via Bonomea 265, 34136 Trieste, Italy, and INFN, Sezione di
Trieste, Italy; } 

$^{(b)}$ {\it Departamento de F\'{i}sica, Universidade Estadual deLondrina, Caixa Postal 6001, Londrina-PR, Brazil}
}


\end{center}

\vspace{0.1in}
\begin{center}
{\bf Abstract}
\end{center}
There is a remarkable connection between quantum generating functions of field theory and formal power series associated with dimensions of chains and homologies of suitable Lie algebras. We discuss the homological aspects of this connection with its applications to partition functions of the minimal three-dimensional gravities in the space-time asymptotic to $AdS_3$, which also describe the three-dimensional Euclidean black holes, the pure $\cN = 1$ supergravity, and a sigma model on $N$-fold generalized symmetric products. We also consider in the same context elliptic genera of some supersymmetric sigma models.
These examples can be considered as a straightforward application of the machinery of  modular forms and spectral functions (with values in the congruence subgroup of $SL(2, {\mathbb Z})$) 
to partition functions represented by means of formal power series that encode Lie algebra properties.



\newpage
\tableofcontents
\newpage

\section{Introduction}

In this paper we start to explore a remarkable connection between quantum partition functions and formal power series associated with dimensions of chains and homologies of Lie algebras
(Euler-Poincar\'{e} formula). 
The combinatorial identities we will be concerned with play  an important role in a number of physical models. In particular, such identities, related to appropriate Lie algebras (and Lie groups), are linked to partition functions of quantum gravity and extended supergravity, elliptic genera of superconformal quantum theory and supersymmetric sigma models, and play a special role in string and black hole dynamics. From a concrete point of view this paper  consists of  applications of modular forms (and spectral functions related to the congruence subgroup of $SL(2, {\mathbb Z})$) to partition functions connected to suitable Lie algebras.
The connection referred to above is particularly striking in the case of the correspondence between
three-dimensional quantum gravity, in a space-time asymptotic to $AdS_{3}$, and $2D$ CFT's. On the $AdS_{3}$
one has Selberg spectral functions and Ruelle functions; on the CFT side partition functions and modular forms.
What we would like to show is that these objects have a common background constituted by Euler-Poincar\'e and Macdonald identities, which, in turn, describe homological aspects of (finite or infinite dimensional) Lie algebra representations.

\noindent
In Sect. \ref{Gravity} we consider theory of the minimal three-dimensional quantum gravity in a space-time asymptotic to $AdS_{3}$, black holes and the Neveu-Schwarz and Ramond sector of $\cN=1$ supergravity. The symmetry group of $AdS_3$ gravity (with appropriate boundary conditions) is generated by the Virasoro algebra \cite{Brown}, and the one-loop partition function is indeed the partition function of a conformal field theory in two dimensions ($CFT_2$) \cite{Giombi}. 

In Sect. \ref{Spectral} we introduce the Petterson-Selberg and Ruelle spectral functions of hyperbolic three-geometry and then, in Sect. \ref{3-Gravity}, we analyze the one-loop corrections to three-dimensional gravity and show that the holomorphic contribution to partition functions corresponds to the formal character of the {\rm Vir}-module. We calculate the asymptotic limit for the coefficient in the expansion of partition functions and conclude that it has a universal form indicating that $q$-series inherit Lie algebra properties. Quantum partition functions for the three-dimensional black holes we compute
in Sect. \ref{BH-Geometry}. We extend our results to $\cN=1$ supergravity in Sect. \ref{N=1}. We show that infinite series of quantum corrections for three-dimensional gravity and the Neveu-Schwarz and Ramond sector of ${\cN}=1$ supergravity 
can be reproduced in terms of Selberg-type spectral functions in a holomorphically factorized theory. 
\\

\noindent
In Sect. \ref{Outline} we discuss the homological aspects of the Macdonald identities which are related to Lie algebras (the Euler-Poincar{\'e} formula). In reproducing the main results we follow the book \cite{Fuks}. This section is designed to provide the reader with a brief introduction to homological aspects of differential complexes and indicate how combinatorial identities, useful for derivation of partition functions, could be obtained from an initial complex of (graded) Lie algebras.
\\

\noindent
In Sect. \ref{N-Folds} we consider examples for which we show that partition functions can be written in terms of Selberg-type (Ruelle) spectral functions associated with $q$-series,  although the hyperbolic side
remains to be explored. In our examples we concentrate on a (supersymmetric) sigma models. 

In Sect. \ref{HH} we explain briefly a relation between the Heisenberg algebra, its representation and the Hilbert scheme of points. 
In Sect. \ref{Sigma} we analyze sigma models on the $N$-fold symmetric product $X^N/{\mathfrak S}_N$ (${\mathfrak S}_N$ is the symmetric group of $N$ elements). The elliptic genus of a supersymmetric sigma model is (almost) an authomorphic form for $O(3,2,{\mathbb Z})$. (For trivial line bundles the elliptic genus degenerates to the Euler number (or Witten index) and can be rewritten in terms of spectral functions of hyperbolic geometry). 
The notion of elliptic genus was introduced in \cite{Ochanine} with applications in \cite{Witten00}.
It has been argued that it is possible to use elliptic modular forms to write generating functions (elliptic genera) of quantum field theory for infinite series of operators associated with homologies of finite dimensional Lie algebras.
The elliptic genus can be interpreted as a natural invariant in generalized cohomology theory, the so-called elliptic cohomology \cite{Landweber1,Landweber2}. Such cohomology could be proposed as a generalization of $K$-theory, and at this point partition functions might be related to elliptic cohomology and $K$-theory. 
We pay attention to the special case of algebraic structures of the $K$-groups $K_{\widetilde{H}\Gamma_N}(X^N)$ of $\Gamma_N$-equivariant Clifford supermodulus on $X^N$, following the lines of \cite{Wang}. This case is important since 
the direct sum $\cF_{\Gamma}^{-}(X)= \oplus_N^{\infty}K_{\widetilde{H}\Gamma_N}(X^N)$ carries naturally a Hopf algebra structure, and is isomorphic to the Fock space of a twisted Heisenberg superalgebra with $K_{\widetilde{H}\Gamma_N}(X)\cong K_{\Gamma}(X)$.
We represent in terms of the Ruelle spectral function the dimension of a direct sum of the equivariant $K$-groups (related to a suitable supersymmetric algebra).
We analyse also elliptic genera for generalized wreath and symmetric products on $N$-folds; these cases are examples of straightforward applications of the machinery of modular forms and spectral functions discussed above.

In Sect. \ref{Orbifold} we study the generating functions on (loop) orbispaces with its relation to the Macdonald polynomials and the Ruelle spectral functions.

In Sect. \ref{K3} we discuss a multiple $D$-brane description which leads to a two-dimensional sigma model with target space the sum of symmetric products of the $K3$ space \cite{Strominger1}. The counting of the BPS states is naturally related to the elliptic genus of this sigma model, while the degeneracies are given in terms of the denominator of a generalized super Kac-Moody algebra which admits description in terms of spectral functions. 
\\

\noindent
Finally in Sect. \ref{Conclusions} (Conclusions) we briefly outline issues and futher perspectives for analysis of gravity partition functions, string $N$-fold solutions and  deformation quantization.

\section{Three-dimensional gravity, black holes and $\cN=1$  supergravity}
\label{Gravity}

In this section we would like to shed light on some aspects of the
$AdS_3/CFT_2$ correspondence.
It is known that the geometric structure of three-dimensional gravity (and black holes) allows for exact computations since its Euclidean counterpart is locally isomorphic to the constant curvature hyperbolic space.
Because of the $AdS_3/CFT_2$ correspondence, we expect a correspondence between spectral functions related to Euclidean $AdS_3$ and modular-like functions (Poincar\'e series)
\footnote{ 
The modular forms in question are the forms for the congruence subgroup of $SL(2, {\mathbb Z})$,
which is viewed as the group that leaves fixed one of the three non-trivial spin structures on an elliptic curve.}.
We assume that this correspondence occur when the arguments of spectral functions take values on a Riemann surface, viewed as the conformal boundary of $AdS_3$. According to the holographic principle, there exist strong ties between certain field theory quantities on 
the bulk of an $AdS_3$ manifold and related quantities on its boundary at infinity. To be more precise, the classes of Euclidean $AdS_3$ spaces are quotients of the real hyperbolic space by a discrete group (a Schottky group). The boundary of these spaces can be compact oriented surfaces with conformal structure (compact complex algebraic curves). A general formulation of the {\it Holography Principle} states that there is a correspondence between a certain class of fields, their properties and their correlators in the bulk space, where gravity propagates, and a class of primary fields, with their properties and correlators of conformal theory on the boundary. More precisely, the set of scattering poles in $3D$ coincides with the zeroes of a Selberg-type spectral function \cite{Perry,Bytsenko4} {(see Sect. \ref{Spectral} below)}; thus encoded in a Selberg function is the spectrum of a three-dimensional model.

\begin{itemize}
\item In the framework of this general principle we would like to illustrate the correspondence between spectral functions of hyperbolic three-geometry (its spectrum being encoded in the Petterson-Selberg spectral functions, see Remark \ref{remark1} below) and Poincar\'e series associated with conformal structure in two dimensions. 
\end{itemize}

\subsection{Spectral functions of hyperbolic three-geometry}
\label{Spectral}

The Euclidean sector of $AdS_3$ has an orbifold description $H^3/\Gamma$. The complex unimodular group $G=SL(2, {\mathbb C})$
acts on the real hyperbolic three-space $H^3$ in a standard way, namely for $(x,y,z)\in H^3$ and $g\in G$, one gets
$g\cdot(x,y,z)= (u,v,w)\in H^3$. Thus for $r=x+iy$,\,
$g= \left[ \begin{array}{cc} a & b \\ c & d \end{array} \right]$,
$
u+iv = [(ar+b)\overline{(cr+d)}+ a\overline{c}z^2]\cdot
[|cr+d|^2 + |c|^2z^2]^{-1},\,
w = z\cdot[
{|cr+d|^2 + |c|^2z^2}]^{-1}\,.
$
Here the bar denotes the complex conjugation. Let $\Gamma \in G$ be the discrete group of $G$
defined as 
\begin{eqnarray}
\Gamma & = & \{{\rm diag}(e^{2n\pi ({\rm Im}\,\tau + i{\rm Re}\,\tau)},\,\,  e^{-2n\pi ({\rm Im}\,\tau + i{\rm Re}\,\tau)}):
n\in {\mathbb Z}\}
= \{{\mathfrak g}^n:\, n\in {\mathbb Z}\}\,,
\nonumber \\
{\mathfrak g} & = &
{\rm diag}(e^{2\pi ({\rm Im}\,\tau + i{\rm Re}\,\tau)},\,\,  e^{-2\pi ({\rm Im}\,\tau + i{\rm Re}\,\tau)})\,.
\end{eqnarray}
One can define the Selberg-type zeta function for the group 
$\Gamma = \{{\mathfrak g}^n : n \in {\mathbb Z}\}$ generated by a single hyperbolic element of the form ${\mathfrak g} = {\rm diag}(e^z, e^{-z})$, where $z=\alpha+i\beta$ for $\alpha,\beta >0$. In fact we will take
$\alpha = 2\pi {\rm Im}\,\tau$, $\beta= 2\pi {\rm Re}\,\tau$. For the standard action of $SL(2, {\mathbb C})$ on $H^3$ one has
\begin{equation}
{\mathfrak g}
\left[ \begin{array}{c} x \\ y\\ z \end{array} \right]
=
\left[\begin{array}{ccc} e^{\alpha} & 0 & 0\\ 0 & e^{\alpha} & 0\\ 0
& 0 & \,\,e^{\alpha} \end{array} \right]
\left[\begin{array}{ccc} \cos(\beta) & -\sin (\beta) & 0\\ 
\sin (\beta) & \,\,\,\,\cos (\beta) & 0
\\ 0 & 0 & 1 \end{array} \right]
\left[\begin{array}{c} x \\ y\\ z \end{array} \right]
\,.
\end{equation}
Therefore ${\mathfrak g}$ is the composition of a rotation in ${\mathbb R}^2$ with complex eigenvalues $\exp (\pm i\beta)$ and a dilatation $\exp (\alpha)$. The Petterson-Selberg spectral function $Z_\Gamma (s)$ can be attached to ${H}^3/\Gamma$ (see also \cite{Bytsenko1}) as follows:
\begin{equation}
Z_\Gamma(s) :=\prod_{\stackrel{k_1,k_2\geq
0}{k_1,k_2\in\mathbb{Z}}}^\infty[1-(e^{i\beta})^{k_1}(e^{-i\beta})^{k_2}e^{-(k_1+k_2+s)\alpha}]\,.
\label{zeta00}
\end{equation}
Zeros of $Z_\Gamma (s)$ are precisely the complex numbers
$
\zeta_{n,k_{1},k_{2}} = -\left(k_{1}+k_{2}\right)+i\left(k_{1}-
k_{2}\right)\beta/\alpha+ 2\pi  in/\alpha$\,, 
{\rm (}$n \in {\mathbb Z}${\rm )}. 
The logarithm of $Z_\Gamma (s)$ for ${\rm Re}\, s> 0$ is given by 
{\rm \cite{Bytsenko1}}
\begin{equation}
{\rm log}\, Z_{\Gamma} (s)   =  
-\frac{1}{4}\sum_{n\in {\mathbb Z}_+}^{\infty}\frac{e^{-n\alpha(s-1)}}
{n[\sinh^2\left(\frac{\alpha n}{2}\right)
+\sin^2\left(\frac{\beta n}{2}\right)]}\,.
\label{logZ}
\end{equation}

In order to appreciate the importance of zeros of $Z_\Gamma (s)$
it is necessary to briefly outline the spectral
analysis of the Laplacian $\Delta_{\Gamma} = z^2 (\partial^2/\partial x^2+ \partial^2/\partial y^2 +\partial^2\partial z^2)-z\partial/\partial z $, where a key notion is played by the scattering resonances (see for more detail \cite{Perry}).
The space of square-integrable functions on the black hole
${H}^3/\Gamma $, with respect to the Riemannian volume element $dV$, has an orthogonal decomposition
\begin{equation}
L^2\left({H}^3/\Gamma, dV\right)=\sum_{m,n\in\mathbb{Z}}\oplus H_{mn} \,\,\,
{\rm with}\,\,\, {\rm Hilbert}\,\,\, {\rm space}
\,\,\, {\rm isomorphisms}\,\,\, H_{mn}\cong
L^2\left(\mathbb{R}_{+},dt\right)\,,
\end{equation}
where $\mathbb{R}_{+}$ is the space of positive real numbers.
A spectral decomposition takes the form
\begin{equation}
-\Delta_\Gamma\simeq\sum_{m,n\in\mathbb{Z}}\oplus L_{mn}
\,,\,\,\,\,\,\,\,
L_{mn}=-\frac{d^2}{dt^2}+1+V_{mn}(t)\,,
\label{L}
\end{equation}
where $L_{mn}$ are the Schr\"{o}dinger operators with P\"{o}schel-Teller potentials
$
V_{mn}(t)= (k_{mn}^2+ 1/2){\rm sech}^2t+ (m^2- 1/4){\rm cosh}^2t$,\,
$k_{mn}:= - m\beta/\alpha+ \pi n/\alpha.
$
(For details of this and the following remarks the reader can
consult \cite{Perry}, for example.)
The Schr\"{o}dinger equation
$
\Psi ''(x)+[E-V_{mn}(x)]\Psi(x)=0,
$
which reduces to the eigenvalue problem $L_{mn}\Psi=k^2\Psi$ for
$E=k^2-1$, has a known solution $\Psi^+(x)$ (in terms of the
hypergeometric function) with asymptotics
\begin{equation}
\Psi^+(x)\sim
\frac{e^{ikx}}{T_{mn}(k)}+
\frac{R_{mn}(k)}{T_{mn}(k)}e^{-ikx},
\end{equation}
for
{\it reflection} and {\it transmission} coefficients
$T_{mn}(k)$, $R_{mn}(k)$ respectively.
Define $k \stackrel{def}{=} i(1-s)$. Then one can form the scattering matrix
$
\left[{\mathfrak S}_{mn}(s)\right]\stackrel{def}{=}
\left[R_{mn}(k)\right]
$
of $-\Delta_{\Gamma}$, whose entries are quotients of gamma
functions with "trivial poles" $s=1+j$,
$j\in {\mathbb Z}_+\cup \{0\},$ and non-trivial poles
\begin{equation}
s_{mnj}^{\pm}:= -2j-\left|m\right|\pm i\left|k_{mn}\right|.
\label{poles}
\end{equation}
The $s^{\pm}_{mnj}$ are the scattering resonances.
\begin{remark} \label{remark1}
It is remarkable fact that the set of scattering poles in 
{\rm (}\ref{poles}{\rm )} coincides with the zeros of $\zeta_{n,k_{1},k_{2}}$,
as it can be verified. Thus encoded in $Z_{\Gamma}(s)$ is the spectrum of a three-dimensional model.
\end{remark}

\noindent
{\bf Generating and spectral functions.}
Using the equality  
$
\sinh^2\left(\alpha n/2\right)
+ \sin^2\left(\beta n/2\right)$ $ = |\sin(n\pi \tau)|^2 =
|1-q^n|^2/(4|q|^{n})
$
and Eq. (\ref{logZ}) we get
\begin{eqnarray}
{\rm log}\prod_{m=\ell}^{\infty}(1- q^{m+\varepsilon}) 
& = & 
\sum_{m=\ell}^{\infty} 
{\rm log}(1 - q^{m+\varepsilon}) 
= 
-\sum_{n=1}^{\infty}
\frac{q^{(\ell + \varepsilon)n}
(1- \overline{q}^{n})|q|^{-n}}
{4n|\sin(n\pi \tau)|^2} 
\nonumber \\
&=&
{\rm log}\,\left[\frac{Z_\Gamma(\xi (1-it))}{Z_\Gamma(
\xi (1-it)+1+it)}\right]\,,
\label{modular1}
\end{eqnarray}
\begin{eqnarray}
{\rm log}\prod_{m=\ell}^{\infty}(1- \overline{q}^{m+\varepsilon}) 
& = & 
\sum_{m=\ell}^{\infty} 
{\rm log}(1 - {\overline q}^{m+\varepsilon}) 
= 
-\sum_{n=1}^{\infty}
\frac{{\overline q}^{(\ell + \varepsilon)n}
(1- {q}^{n})|q|^{-n}}
{4n|\sin(n\pi \tau)|^2} 
\nonumber \\
& = &
{\rm log}\,\left[\frac{Z_\Gamma(\xi (1+it))}{Z_\Gamma(
\xi (1+it)+1-it)}\right]\,,
\label{modular2}
\end{eqnarray}
\begin{eqnarray}
{\rm log}\prod_{m=\ell}^{\infty}(1+ {q}^{m+\varepsilon}) 
& = & 
\sum_{m=\ell}^{\infty} 
{\rm log}(1 + {q}^{m+\varepsilon}) 
= 
-\sum_{n=1}^{\infty}
\frac{(-1)^n q^{(\ell + \varepsilon)n}
(1- {\overline q}^{n})|q|^{-n}}
{4n|\sin(n\pi \tau)|^2} 
\nonumber \\
& = &
{\rm log}\,\left[\frac{Z_\Gamma(\xi (1-it) + 
i\eta(\tau))}{Z_\Gamma(\xi (1+it)+1-it + i\eta(\tau))}\right]\,,
\label{modular3}
\end{eqnarray}
\begin{eqnarray}
{\rm log}\prod_{m=\ell}^{\infty}(1+ {\overline q}^{m+\varepsilon}) 
& = & 
\sum_{m=\ell}^{\infty} 
{\rm log}(1 + {\overline q}^{m+\varepsilon}) 
= 
-\sum_{n=1}^{\infty}
\frac{(-1)^nq^{(\ell + \varepsilon)n}
(1- {q}^{n})|q|^{-n}}
{4n|\sin(n\pi \tau)|^2} 
\nonumber \\
& = &
{\rm log}\,\left[\frac{Z_\Gamma(\xi(1+it) + i\eta(\tau))}{Z_\Gamma(\xi(1+it)+1-it + i\eta(\tau))}\right]\,,
\label{modular4}
\end{eqnarray}
where $\ell \in {\mathbb Z}_+,\, \varepsilon \in {\mathbb C}$, $t = {\rm Re}\,\tau/{\rm Im}\,\tau$, $\xi = \ell + \varepsilon$ and $\eta (\tau)= \pm(2\tau)^{-1}$. 

Let us next introduce some well--known functions and their modular properties under the action of $SL(2, {\mathbb Z})$. The special cases associated with (\ref{modular1}), (\ref{modular2})
are (see for details \cite{Kac}):
\begin{eqnarray}
f_1(q) & = & q^{-\frac{1}{48}}\prod_{m\in {\mathbb Z}_+}
(1-q^{m+\frac{1}{2}})\,\, = \,\, \frac{\eta_D(q^{\frac{1}{2}})}{\eta_D(q)}\,,
\\
f_2(q) & = & q^{-\frac{1}{48}}\prod_{m\in {\mathbb Z}_+}
(1+q^{m+\frac{1}{2}})\,\, = \,\, \frac{\eta_D(q)^2}{\eta_D(q^{\frac{1}{2}})\eta_D(q^2)}\,,
\\
f_3(q) & = & \,\, \,\, q^{\frac{1}{24}}\prod_{m\in {\mathbb Z}_+}
(1+q^{m+1})\,\, = \,\, \frac{\eta_D(q^2)}{\eta_D(q)}\,,
\end{eqnarray}
where 
$
\eta_D(q) \equiv q^{1/24}\prod_{n\in {\mathbb Z}_+}(1-q^{n})
$
is the Dedekind $\eta$-function. The linear span of $f_1(q), f_2(q)$
and $f_3(q)$ is $SL(2, {\mathbb Z})$-invariant \cite{Kac}
(
$
\!g\in \left[\begin{array}{cc} a & b \\ c & d \end{array}\right],\,
$
$
g \cdot f(\tau) = f\left(\frac{a\tau + b}{c\tau + d}\right)
$
\!).
It can be shown that these are modular forms of weight 0 for the principal congruence subgroup $\Gamma^{48}$ (see also $H_\bullet(K3; {\mathbb Z})$ in the integral homology of $K3$, Sect. \ref{K3}).

For a closed oriented hyperbolic three-manifolds of the form $X= H^3/\Gamma$ (and any acyclic orthogonal representation of $\pi_1(X)$) the analytic torsion takes the form {\rm \cite{Fried}}
\footnote{
The vanishing theorems for type $(0, q)$ cohomology $H^q(X; E)$
of locally symmetric spaces $X$ have been formulated in \cite{Floyd,Abdalla}.
}: 
$[T_{\rm an}(X)]^2 = \cR(0)$, where $\cR(s)$ is the Ruelle function. 
A Ruelle function $\cR(s)$ can be defined for large 
${\rm Re}\,s$ and continued meromorphically to the entire complex plain $\mathbb C$ {\rm \cite{Deitmar}}. $\cR(s)$ is an alternating product of more complicate factors, each of which is a Selberg spectral function, 
\begin{equation}
\cR(s):= \prod_{p=0}^{{\rm dim}X-1}Z_{\Gamma}(s+p)^{(-1)^p}.
\end{equation}

Let us introduce next the Ruelle functions $\cR(s),\, \cR(\overline{s}),\, \cR(\sigma),\, \cR(\overline{\sigma})$:
\begin{eqnarray}
\prod_{n=\ell}^{\infty}(1- q^{n+\varepsilon}) 
& = & \prod_{p=0, 1}Z_\Gamma(s+p(1+it))^{(-1)^p} =
\cR(s=\xi(1-it)),
\\
\prod_{n=\ell}^{\infty}(1- \overline{q}^{n+\varepsilon}) 
& = & \prod_{p=0, 1}Z_\Gamma(\overline{s}+p(1-it))^{(-1)^p} =
\cR(\overline{s}=\xi(1+it)),
\\
\prod_{n=\ell}^{\infty}(1+ q^{n+\varepsilon}) 
& = & \prod_{p=0, 1}Z_\Gamma(\sigma+p(1+it))^{(-1)^p} =
\cR(\sigma =\xi(1-it) + i\eta(\tau)),
\\
\prod_{n=\ell}^{\infty}(1+ \overline{q}^{n+\varepsilon}) 
& = & \prod_{p=0, 1}Z_\Gamma(\overline{\sigma}+p(1-it))^{(-1)^p} =
\cR(\overline{\sigma}=\xi(1+it) + i\eta(\tau))\,.
\end{eqnarray}
{ Since
$
f_1(q)\cdot f_2(q)\cdot f_3(q) = 1
$
we get
\begin{equation}
\cR(s=3/2-(3/2)it)\cdot \cR(\sigma =3/2-(3/2)it+i\eta(\tau)) 
\cdot \cR(\sigma =2-2it+i\eta(\tau)) = 1\,.
\end{equation}
}
A set of useful generating functions is collected in Table {\rm 
\ref{Table1}}.
\begin{table}\label{Table1}
\begin{center}
\begin{tabular}
{l l }
Table {}  \ref{Table1}. \,\,\,
List of generating functions
\\
\\
\hline
\\
$ \prod_{n=\ell}^{\infty}(1-q^{n+\varepsilon})=
\left[\frac{Z_\Gamma(\xi (1-it))}
{Z_\Gamma(\xi (1 - it)+ 1+ it)}\right]\,\,\,\,\,\,\,\,\,\,\,\,
= \,\, \cR(s= \xi(1-it))$ 
\\
\\
$ \prod_{n=\ell}^{\infty}(1- \overline{q}^{n+\varepsilon})=
\left[\frac{Z_\Gamma(\xi (1+it))}
{Z_\Gamma(\xi (1 + it)+ 1- it)}\right]\,\,\,\,\,\,\,\,\,\,\,\,
= \,\, \cR(\overline{s}= \xi(1+it))$ 
\\
\\
$
\prod_{n=\ell}^{\infty}(1+q^{n+\varepsilon})=
\left[\frac{Z_\Gamma(\xi(1-it) + i\eta(\tau))}
{Z_\Gamma(\xi (1-it) + i\eta(\tau)+ 1+it)}\right]
= \,\, \cR(\sigma = \xi(1-it) + i\eta(\tau))
$ 
\\
\\
$
\prod_{n=\ell}^{\infty}(1+ \overline{q}^{n+\varepsilon})=
\left[\frac{Z_\Gamma(\xi(1+it) + i\eta(\tau))}
{Z_\Gamma(\xi (1+it) + i\eta(\tau)+ 1-it)}\right]
= \,\, \cR(\overline{\sigma} = \xi(1+it) + i\eta(\tau))
$ 
\\
\\
$ \prod_{n=2}^{\infty}(1-q^n)\,\,\,\,\, = 
\left[\frac{Z_\Gamma(2-2it)}
{Z_\Gamma(3-it)}\right] \,\,\,\,\,\,\,\,\,\,\,\,\,\,\,\,\,\,\,\,
\,\,\,\,\,\,\, = \,\,\cR(s= 2-2it)$ 
\\
\\
$ \prod_{n=2}^{\infty}(1-\overline{q}^n) \,\,\,\,\, = 
\left[\frac{Z_\Gamma(2+2it)}
{Z_\Gamma(3+it)}\right] \,\,\,\,\,\,\,\,\,\,\,\,\,\,\,\,\,\,\,\,
\,\,\,\,\,\,\, = \,\, \cR(\overline{s}=2+2it)$
\\
\\
$ \prod_{n=2}^{\infty}(1+q^n) \,\,\,\,\, = 
\left[\frac{Z_\Gamma(2-2it + i\eta(\tau)}
{Z_\Gamma(3-it + i\eta(\tau)}\right]\,\,\,\,\,\,\,\,
\,\,\,\,\,\,\, = \,\, \cR(\sigma = 2-2it + i\eta(\tau))$
\\
\\
$ \prod_{n=2}^{\infty}(1+\overline{q}^n) \,\,\,\,\, = 
\left[\frac{Z_\Gamma(2+2it + i\eta(\tau))}
{Z_\Gamma(3+it + i\eta(\tau))}\right] \,\,\,\,\,\,\,\,\,
\,\,\,\,\, = \,\, \cR(\overline{\sigma}= 2+2it + i\eta(\tau))$
\\
\\
$ \prod_{n=2}^{\infty}(1-q^{n-\frac{1}{2}}) =
\left[\frac{Z_\Gamma(\frac{3}{2}-\frac{3}{2}it)}
{Z_\Gamma(\frac{5}{2}-\frac{1}{2}it)}\right]
\,\,\,\,\,\,\,\,\,\,\,\,\,\,\,\,\,\,\,\,\,\,\,
= \,\,\cR(s= 3/2- (3/2)it)$
\\
\\
$ \prod_{n=2}^{\infty}(1-\overline{q}^{n-\frac{1}{2}}) =
\left[\frac{Z_\Gamma(\frac{3}{2}+\frac{3}{2}it)}
{Z_\Gamma(\frac{5}{2}+\frac{1}{2}it)}\right]\,\,\,\,\,\,\,
\,\,\,\,\,\,\,\,\,\,\,\,\,\,\,\, 
= \,\,\cR(\overline{s}= 3/2+ (3/2)it)$
\\
\\
$ \prod_{n=2}^{\infty}(1+ q^{n-\frac{1}{2}}) =
\left[\frac{Z_\Gamma(\frac{3}{2}-\frac{3}{2}it + i\eta(\tau))}
{Z_\Gamma(\frac{5}{2}-\frac{1}{2}it + i\eta(\tau))}\right]
\,\,\,\,\,\,\,\,\,\,
= \,\,\cR(\sigma = 3/2 - (3/2)it + i\eta(\tau))$
\\
\\
$ \prod_{n=2}^{\infty}(1+ \overline{q}^{n-\frac{1}{2}}) =
\left[\frac{Z_\Gamma(\frac{3}{2}+\frac{3}{2}it + i\eta(\tau))}
{Z_\Gamma(\frac{5}{2}+\frac{1}{2}it + i\eta(\tau))}\right]
\,\,\,\,\,\,\,\,\,\,
= \,\, \cR(\overline{\sigma}= 3/2+ (3/2)it + i\eta(\tau))$
\\
\\
\hline
\end{tabular}
\end{center}
\end{table}

\noindent
{\bf Asymptotic expansions.}
The coefficients in the expansion of these generating
functions in its final form are not always known.
An interesting result has been found in \cite{Perchik}: in the infinite product 
\begin{equation}
\prod_{{\stackrel{-1\leq a\leq \infty}{-a-2\leq b\leq a+2}}}(1-t^bx^a), \,\,\,\,\,\,\, {\rm where}\,\,\,\,\, b\equiv a \,{\rm mod}\,2
\,\,\,\,\, {\rm and} \,\,\,\,\, 
(a, b)\neq (0, 0)\,,
\end{equation} 
the coefficient of $x^{2i} (= t^0x^{2i})$ is equal to $2A_i = 2\sum_{q\geq 0}(-1)^q{\rm dim}\,H^{[i]}_q(H; {\mathfrak g}{\mathfrak l}(2, {\mathbb C}))$. Here the $H$ is a Lie algebra which can be described by means of its central extension, the Poisson algebra. $H$ has a natural $({\mathbb Z}\oplus{\mathbb Z})$-grading and therefore the cited homologies become graded too (if $i\neq j$ then $H^{(i, j)} =0$, and $H^{(i, i)}\equiv H^{[i]}$).  We shall simplify the calculations and compute the asymptotic limit of the above coefficient. 
\begin{eqnarray}
\prod_{n= \ell}^\infty
\left(1- q^{n+\varepsilon}\right)^{-d_n} & = &
\prod_{n= 1}^\infty \left(1- q^{n+\xi -1}\right)^{-d_n}
= 1 + \sum_{N = 1}^{\infty} \cB(N)q^{N + \xi -1}\,,
\label{BN1}
\\
\prod_{n= \ell}^\infty \left(1- \overline{q}^{n+\varepsilon}\right)^{-d_n} & = &
\prod_{n= 1}^\infty\left(1-
\overline{q}^{n+\xi -1}\right)^{-d_n}
= 
1 + \sum_{N = 1}^{\infty} \overline{\cB}(N)\overline{q}^{N + \xi -1}\,,
\label{BN2}
\\
\prod_{n= \ell}^\infty\left(1+
q^{n+\varepsilon}\right)^{-d_n} & = &
\prod_{n= 1}^\infty\left(1-
q^{n+\xi + \eta(\tau)-1}\right)^{-d_n}\,,
\label{BN3} 
\\
\prod_{n= \ell}^\infty\left(1+
\overline{q}^{n+\varepsilon}\right)^{-d_n} & = &
\prod_{n= 1}^\infty\left(1-
\overline{q}^{n+\xi + \eta(\tau) -1}\right)^{-d_n}\,,
\label{BN4}
\end{eqnarray}
where as before $\exp (2\pi i\tau) = 
\exp (-2\pi{\rm Im}\,\tau + 2\pi i {\rm Re}\,\tau)$,\, 
$\varepsilon \geq 0$, $\eta(\tau) = \pm(2\tau)^{-1}$, $q^{\eta(\tau)}\equiv \exp (\pm \pi i) = -1$, and $d_n \equiv {\rm dim}\,{\mathfrak g}_n \,
({\rm or\,\, rank}\,{\mathfrak g}_n) > 0$. Let us consider Eq. (\ref{BN1}) (Eqs. (\ref{BN2})-(\ref{BN4}) can be treated similarly). Suppose that ${\rm Im}\,\tau >0$\, ($|q|< 1$) and let
\begin{equation}
{\mathfrak D}(s; \xi)= \sum_{N=1}^{\infty}d_N (N+\xi -1)^{-s} \,,\,\,\,\,\,\,\,\, s= \sigma +i\varrho,
\end{equation} 
be the associated Dirichlet series which converges for $0<\sigma<p$.
Assume that ${\mathfrak D}(s;\varepsilon)$ can be analytically continued in the region $\sigma\geq -C_0\, (0< C_0 < 1)$ and here ${\mathfrak D}(s; \varepsilon)$ is analytic except for a pole of order one at $s=p$ with residue $A(p)$. We assume also that ${\mathfrak D}(s;\varepsilon) = O(|\varrho|^{C_1})$ uniformly at $|\varrho|\rightarrow \infty$, where $C_1$ is a fixed positive real number. 
The Mellin-Barnes representation leads to the formula
\begin{equation}
\sum_{n=1}^{\infty}{\rm log}(1-q^{n+\xi -1})^{-d_n} =
\int_{1+p-i\infty}^{1+p+i\infty} (2\pi i)^{-s-1}\tau^{-s}
\zeta_R(s+1)\Gamma(s){\mathfrak D}(s; \xi) ds\,.
\label{integral} 
\end{equation}
The integral (\ref{integral}) has a first order pole at $s=p$ and a second order pole at $s=0$. One can shift the vertical contour of intgration from $2\pi {\rm Im}\,\tau = 1+p$ to $2\pi {\rm Im}\,\tau = - C_0$ and make use of the theorem of residues:
\begin{eqnarray}
\sum_{n=1}^{\infty}{\rm log}(1-q^{n+\varepsilon})^{-d_n} & = &
A(p) \Gamma(p)\zeta_R(1+p)(2\pi i\tau)^{-p} + 
((d/ds) - 1)\,{\mathfrak D}(0; \xi) 
\nonumber \\
& + & 
\int_{-C_0-i\infty}^{-C_0 + i\infty} (2\pi i)^{-s-1}\tau^{-s}
\zeta_R(s+1)\Gamma(s){\mathfrak D}(s; \xi) ds\,.
\label{integral2}
\end{eqnarray}
The absolute value of the integral in (\ref{integral2}) can be estimated to behave as $O((2\pi {\rm Im}\,\tau)^{C_0})$, and therefore the expansion
\begin{eqnarray}
\!\!\!\!\!
\sum_{n=1}^{\infty}{\rm log}(1-q^{n+\xi -1})^{-d_n} & = &
\exp \{A(p)\Gamma(p)\zeta_R(1+p)(2\pi {\rm Im}\,\tau)^{-p}
\nonumber \\
& - & 
{\mathfrak D}(0; \xi) +  
(d/ds){\mathfrak D}(0;\xi) + O((2\pi{\rm Im}\,\tau)^{C_0}) \}\,,
\\
\!\!\!\!\!
\sum_{n=1}^{\infty}{\rm log}(1-q^{n+\xi -1})^{-d_n} & = &
O (\exp \{A(p)\Gamma(p)\zeta_R(1+p)(2\pi {\rm Im}\,\tau)^{-p}
- C (2\pi {\rm Im}\,\tau)^{-r}\})\,,
\end{eqnarray}
exists uniformly in ${\rm Re}\,\tau$ as ${\rm Im}\,\tau \rightarrow  0$, provided $|{\rm arg} (-2\pi i \tau)|\leq \pi/4$ and $|{\rm Re}\,\tau|\leq 1/2$. Moreover, the number $r$ is a positive, $C$ being a fixed real number,
$(2\pi{\rm Im}\,\tau)^\nu \leq |{\rm Re}\,\tau|\leq 1/2$, and $\nu = 1+ p/2 -p\delta/4,\,  0< \delta< 2/3$.
By means of the expansion of $\cB(N)$ one arrives at a complete asymptotic limit (the Meinardus result, \cite{Andrews,ElizaldeBook,BytsenkoPhysRep})
\begin{eqnarray}
\cB(N)_{N\rightarrow \infty} & = &
\cC(p)N^{\frac{2{\mathfrak D}(0;\varepsilon)-p-2}{2(1+p)}}
\exp\left\{\frac{1+p}{p}[A(p)
\Gamma(1+p)\zeta_R(1+p)]^
{\frac{1}{1+p}}N^{\frac{p}{1+p}}\right\} 
\nonumber \\
& \times &
[1+{O}(N^{-{\kappa}})]\,,
\label{B(N)}
\\
{\cC}(p) & = & [A(p) \Gamma(1+p)
\zeta_R (1+p)]^{\frac{1-2{\mathfrak D}(0;\varepsilon)}{2p+2}}
\cdot
\frac{\exp\left[(d/ds){\mathfrak D}(0;\varepsilon)\right]}
{[2\pi(1+p)]^{1/2}}\,,
\label{C(p)}
\end{eqnarray}
where
$ 
{\kappa}  = p/(1+p)\cdot {\rm min} (C_0/p - \delta/4,
1/2-\delta).
$

\subsection{Quantum gravity in $AdS_3$}
\label{3-Gravity}
In this section we study generating functions for three-dimensional gravity. It has been shown that the contribution to the
partition function of gravity in a space-time asymptotic to
$AdS_3$ comes from smooth geometries $X = AdS_3/\Gamma$, where $\Gamma$ is a discrete subgroup of $SO(3,1)$. 
To be more precise, it comes from geometries $X_{c,d}$\, (see for detail \cite{Maloney}), where $c$ and $d$ are a pair of relatively prime integers, $c\geq 0$, and
a pair $(c,d)$ identified with $(-c,-d)$. The manifolds
$X_{c,d}$ are all diffeomorphic to each other, and therefore the contribution ${W}_{c,d}(\tau)$ to the partition function can be expressed in terms of any one of them, say ${W}_{0,1}(\tau)$, by a modular transformation. One has the following formula: 
\begin{equation}
{W}_{c,d}(\tau, \overline{\tau}) =  {W}_{0,1}((a\tau+b)/(c\tau+d), \overline{\tau})\,,
\,\,\,\,\,\,\,\,\,\,\,
{W}_{0,1}(\tau, \overline{\tau}):=  |q \bar q|^{-k} \prod_{n=2}^\infty|1-q^n|^{-2}\,.
\label{PF}
\end{equation} 
In Eq. (\ref{PF}) $24k= c_L=c_R= c$, and $c$ is the central charge of a conformal field theory, $q=\exp(2\pi i\tau)=\exp[2\pi(-{\rm Im}\tau +i{\rm Re}\tau)]$ such that 
$|q \overline{q} |^{-k}=\exp(4\pi k {\rm Im}\tau)$ corresponds to
the classical prefactor. 
\begin{comment}
Recall that $H^3/\Gamma$ is also the geometry of a Euclidean three-dimensional black hole. To make correspondence between models one must set $k= (8\pi G)^{-1}$:
\begin{equation}
-\, {\rm log}{W}_{0,1({\rm classical})}(\tau, \overline{\tau})= 
k{\rm log}|\overline{q} q| = 2\pi r_+(4G)^{-1}\,,
\end{equation}
where $r_{+}> 0$ is the outer horizon of a black hole.
This result is the classical part of the contribution; 
Eq. {\rm (\ref{PF})} is one-loop exact as has been claimed in  
{\rm \cite{Maloney}}. Note that the one-loop contribution 
{\rm (\ref{PF})} is qualitatively similar to the quantum correction to a three-dimensional black hole {\rm \cite{Bytsenko2,Bytsenko3}}. On a general ground one would expect that the generating function {\rm (\ref{PF})} be connected to the relevant one-loop determinant. Indeed, one-loop determinants can be resummed with the help of the Poisson summation procedure and may give a possibility to realize this connection. {\rm (}Analogous procedure of Poisson summation for regularized Poincar\'e series associated with $W(\tau)$ has been analysed in {\rm \cite{Maloney}, Sect. 3.2.}{\rm )}
\end{comment}

\subsection{Three-dimensional black holes}
\label{BH-Geometry}

It is known that the one-loop corrections to the three-dimensional gravity on $H^3/\Gamma$ are qualitatively similar to the black hole quantum corrections. In the physics literature usually one assumes that the fundamental domain for the action of a discrete group $\Gamma$ has finite volume. On the other hand a three-dimensional black hole has a Euclidean quotient representation $H^3/\Gamma$ for an appropriate $\Gamma$, where the fundamental domain has infinite hyperbolic volume (for the non-spining black hole one can choose $\Gamma$ to be the Abelian group generated by a single hyperbolic element \cite{Perry}). For discrete groups of isometries of the
three-dimensional hyperbolic space with infinite volume fundamental domain (i.e. for Kleinian groups), Selberg-type functions and trace formulas, excluding fundamental domains with cusps, have been considered in \cite{Perry1}, where the results depend also on previous works \cite{Patterson1}, \cite{Patterson2}. Note that matters are difficult in the case of an infinite-volume setting due to the infinite multiplicity of the continuous spectrum and to the absence of a canonical renormalization of the scattering operator which makes it trace-class. However, for a three-dimensional black hole one can by-pass most of the general theory and proceed more directly to define a Selberg function attached to $H^3/\Gamma$ and establish a trace formula which is a version of the Poisson formula for a resonance (see for detail \cite{Perry}).
In fact, there is a special relation between the spectrum and the {\it truncated} heat kernel of the Euclidean black hole with the Petterson-Selberg spectral function \cite{Bytsenko1}.
Thus the quantum partition function of gravity can be given by means of the Petterson-Selberg spectral function (\ref{zeta00}).
One can summarize the preceding discussion by means of the formula 
\begin{eqnarray}
{\rm log}\,{W}_{\rm gravity}^{\rm 1-loop}(\tau, \overline{\tau}) 
& = & 
[ \prod_{n=2}^\infty (1-q^n)\cdot \prod_{n=2}^\infty (1-\overline{q}^n)]^{-1} 
\nonumber \\
& = &
- {\rm log}\left[
\cR(s= 2-2it)\cdot\cR(\overline{s}=2+2it)\right]\,.
\label{W1-loop}
\end{eqnarray}
Note that if we apply the modular transformation $\tau = (\beta + i\alpha)/2\pi \rightarrow -1/\tau$, i.e. 
$
{\rm Re}\,\tau \rightarrow - ({\rm Re}\,\tau)\cdot
|\tau|^{-2},\,
{\rm Im}\,\tau \rightarrow  
({\rm Im}\,\tau)\cdot |\tau|^{-2},\, 
t \rightarrow - t\,,
$
the result (\ref{W1-loop}) preserves the form. Computation of the one-loop partition function of three dimensional gravity gives the one loop correction to a black hole. Therefore, 
$
{W}_{\rm gravity}^{\rm 1-loop}(\tau, \overline{\tau})
={W}_{(\rm spinning\,\,\, black\,\,\,hole)}^{\rm 1-loop}(\alpha, \beta)\,.
$

We would like to clarify the physical significance of the coefficients $\cB(N)$ (\ref{B(N)}). Making use of comparison with Eq. (\ref{W1-loop}), for the one-loop contribution to the partition function of the three-dimensional gravity (and black holes) we find: $d_n = 1,\, \varepsilon = 1$, and therefore ${\mathfrak D}(s; 1)= \zeta_R(s; 1) \equiv \zeta_R(s)$. Thus the sub-leading quantum correction to the microcanonical entropy (see for example \cite{Maloney}) becomes:
\begin{eqnarray}
S(N) & = & {\rm log}\,\cB (N) = \cC_1(p) N^{\frac{p}{p+1}} +
\cC_2(p) {\rm log}\,N + \cC(p) + O (N^{-\kappa})\,,
\label{entropy-final}
\\
\cC_1(p)  & \equiv &   \frac{1+p}{p}[A(p)
\Gamma(1+p)\zeta_R(1+p)]^{\frac{1}{1+p}}\,,
\,\,\,\,\,\,\,\,\,\,
\cC_2(p) \equiv \frac{2{\mathfrak D}(0;\varepsilon)-p-2}{2(1+p)}\,. 
\end{eqnarray}
In Eq. (\ref{entropy-final}) the second term is the logarithmic correction. Typically this term appears when the entropy is computed in the microcanonical ensemble (as opposite to the canonical one)
\cite{Maloney}. The explicit value of the prefactor $\cC(p)$ in the expansion (\ref{B(N)}) gives the constant term in the final Eq. (\ref{entropy-final})

\subsection{Holomorphic factorization for supergravity quantum corrections}
\label{N=1}
In this section we show that quantum corrections for three-dimensional gravity and the Neveu-Schwarz and Ramond  sector of ${\cN}=1$ supergravity can be reproduced in terms of Selberg-type spectral functions in a holomorphically factorized theory. Before plunging into the subject it is opportune to prepare the ground in order to be able to appreciate the role of the Virasoro algebra in this paper.
The Virasoro algebra is a Lie algebra, denoted {\rm Vir}, over $\mathbb C$
with basis $L_n$ ($n\in {\mathbb Z}$), $c$.  The Lie algebra {\rm Vir} is a (universal) central extension of the Lie algebra of holomorphic vector fields on the punctured complex plane having finite Laurent series. For this reason the {\rm Vir}-algebra plays a key role in conformal field theory. The remarkable link between the theory of highest-weight modules over the {\rm Vir}-algebra, conformal field theory and statistical mechanics was discovered in \cite{Belavin1,Belavin2}. Here we recall a few elements of representation theory of the {\rm Vir}-algebra which in fact are very similar to that of Kac-Moody algebras.

Let $M(c, h)\, (c, h \in {\mathbb C})$ be the Verma module over {\rm Vir} (see for example \cite{Kac}). The {\it conformal central charge} $c$ acts on $M(c, h)$ as $cI$. Since $[L_0, L_{-n}] = n L_{-n}$, $L_0$ is diagonalizable on $M(c, h)$ with spectrum $h+ {\mathbb Z}_{+}$ and with eigenspace decomposition
\begin{equation}
M(c, h) =\bigoplus_{j\in {\mathbb Z}_{+}} M(c, h)_{h+j}\,, 
\end{equation}
where $M(c, h)_{h+j}$ is spanned by elements of the basis 
of $M(c, h)$. It follows that
$
W_j = {\rm dim}\, M(c, h)_{h+j}, 
$
where $W_j$, as a function of $j$, is the {\it classical partition function} \cite{Kac}. It means that the Kostant partition function for {\rm Vir}-algebra is the classical partition function. On the other hand the partition functions can be rewritten in the form
\begin{equation}
{\rm Tr}_{M(c, h)}\, q^{L_0} := \sum_{\lambda}q^{\lambda}{\rm dim}\,
M(c, h)_{\lambda}  = q^h\prod_{j=1}^\infty (1-q^j)^{-1}\,.
\label{ch}
\end{equation}
The series ${\rm Tr}_V\,q^{L_0}$ is called the formal character of the 
{\rm Vir}-module $V$ ($V\in {\mathfrak C}$, where $\mathfrak C$ is the category; the morphisms in $\mathfrak C$ are homomortphism of {\rm Vir}-modules).

For three-dimensional gravity in a real hyperbolic space the one-loop generating function, as a product of holomorphic and antiholomorphic functions has the form (\ref{W1-loop}).
The full gravity partition function (\ref{PF}) admits the factorization
$
{W}_{0, 1}(\tau, \overline{\tau})  =  
W(\tau)_{\rm hol}\cdot W(\overline{\tau})_{\rm antihol},
$
where
\begin{equation}
W(\tau)_{\rm hol} =  q^{-k}\prod_{n=1}^{\infty}(1-q^{n+1})^{-1}\,,\,\,\,\,\,\,\,\,\,
W(\overline{\tau})_{\rm antihol} =  \overline{q}^{-k}\prod_{n=1}^{\infty}(1-\overline{q}^{n+1})^{-1}\,.
\label{holomorphic}
\end{equation}
Note that the holomorphic contribution in (\ref{holomorphic}) corresponds to the formal character of the {\rm Vir}-module (\ref{ch}).

The modulus of a Riemann surface $\Sigma$ of genus one (the conformal boundary of AdS$_3$) is defined up to $\gamma \cdot\tau = (a\tau+ b)/(c\tau + d)$ with 
$\gamma \in SL(2,{\mathbb Z})$.
The generating function as the sum of known contributions 
of states of left- and right-moving modes in the conformal field theory
takes the form \cite{Maloney}
\begin{equation}
\sum_{c,d}{W}_{c,d}(\tau, \overline{\tau})=
\sum_{c,d}{W}_{0,1}((a\tau+b)/(c\tau+d), \overline{\tau})\,.
\label{Z02}
\end{equation}  
We would like to comment about the sum over geometries.
The generating function, including the contribution from the Brown-Henneaux excitations, have the form (useful generating functions are listed in Table \ref{Table1})
\begin{eqnarray} 
\!\!\!\!
\sum_{c,d} {W}_{c, d}(\gamma\cdot\tau, \overline{\tau}) & = &
\sum_{c,d}
\left|q^{-k}
\prod_{n=2}^{\infty}(1-q^{n})^{-1}\right|_{\gamma}^{2}
\nonumber \\
& = & 
\sum_{c,d} \left\{
|q\overline{q}|^{-k}\cdot
[\cR(s=2-2it)]_{\rm hol}^{-1}\cdot 
[\cR(\overline{s}=2+2it)]_{\rm antihol}^{-1}
\right\}_{\gamma}.
\label{summand}
\end{eqnarray}
Here $| ... |_{\gamma}$ denotes the transform of an expression $| ... |$ by $\gamma$. The summand in (\ref{summand}) is independent of the choice of $a$ and $b$ in $\gamma$.
Note that the sum over $c$ and $d$ in (\ref{summand}) should be thought of as a sum over the coset $PSL(2,\mathbb Z)/{\mathbb Z} \equiv (SL(2,\mathbb Z)/\{\pm 1\})/\mathbb Z$.  
Since $ad-bc =1$, we get $(a \tau + b)/(c \tau + d) = a/c\,\, -1/c(c\tau +d)$, one may show that
\begin{equation}
{\rm Im}\,(\gamma \cdot \tau)=\frac{{\rm Im}\,\tau}
{|c\tau+d|^2}\,,\,\,\,\,\,\,\,
{\rm Re}\,(\gamma \cdot \tau)= \frac{a}{c}-\frac{c{\rm Re}\,\tau+d}{c|c\tau+d|^2}\,.
\end{equation}
\\

\noindent
{\bf The Neveu-Schwarz and Ramond sectors of \,the ${\cN}$=1 supergravity.} 
We shall consider only the basic case of ${\cN}$ =1 supergravity
(see \cite{Maloney} for explanation). 
The symmetry group $SL(2,\mathbb R)\times SL(2,\mathbb R)$
of $AdS_3$ is replaced by $OSp(1|2)\times OSp(1|2)$, where
$OSp(1|2)$ is a supergroup whose bosonic part is
$Sp(2,\mathbb R)=SL(2,\mathbb R)$. The boundary CFT has $(1,1)$ supersymmetry (${\cN}=1$ supersymmetry for both left- and right-movers). There are a few closely related choices of possible partition function,  
$
{\rm Tr}\,\exp(-\beta H-i\theta J)
$
or
$
{\rm Tr}\,(-1)^F\exp(-\beta H-i\theta J)\,.
$
The conserved angular momentum $J=L_0-\bar L_0$ generates a rotation at infinity of the asymptotic $AdS_3$ space-time. The operator $(-1)^F$ is equivalent to $(-1)^{2J}$; states of integer or half-integer $J$ are bosonic or fermionic, respectively.  This property is inherited from the perturbative spectrum of Brown-Henneaux excitations.
The trace  could be computed in either the Neveu-Schwarz (NS) or the Ramond (R) sector. One can compute these partition functions  by summing over three-manifolds $X$ that are locally $AdS_3$ and whose conformal boundary is a Riemann surface $\Sigma$ of genus one. The four possible
partition functions associated with NS or R sectors (with or without an insertion of $(-1)^F$) correspond to the four spin structures on $\Sigma$. An element $g$ of $G=SL(2, {\mathbb R})$ acts on a spin structure by
$
g \cdot
\left[ \begin{array}{c} \mu \\ \nu \end{array} \right]
\rightarrow
\left[ \begin{array}{cc} a & b \\ c & d \end{array} \right]\cdot
\left[ \begin{array}{c} \mu \\ \nu \end{array} \right],
\, \mu, \nu \in (1/2){\mathbb Z}/{\mathbb Z}\,,
$
where the four spin structures on the two-torus $\Sigma$ are represented by the column vector, and $\mu, \nu$ take the values (1/2) for antiperiodic (NS) boundary conditions and 0 for periodic (R) ones.
Taking into account the choice of the spin structure on $\Sigma$, one can sum over choices of $X$ such that the given spin structure on $\Sigma$ does extend over $X$. The NS spin structure on $\Sigma$ is compatible with $X_{0,1}$, and therefore $X_{0,1}$ contributes to traces in the NS sector, not the R sector. 
The partition function of left- and right-moving excitations is $F(q, \overline{q}) = {\rm Tr}_{\rm NS}\exp(-\beta H - i\theta J)$.
Let us also analyze partition functions with other spin structures. If we let $\mu=0$,
$\nu=1/2$, then we get $ G(q, \overline{q})=\Tr_{\rm NS}\,(-1)^F\exp(-\beta H-i\theta J)$.
Thus the contribution to $F(q, \overline{q})$ and 
$G(q, \overline{q})$ associated with $X_{0,1}$ for all spin structures becomes \cite{Maloney}: 
\begin{eqnarray}
F_{0,1}(\tau, \overline{\tau}) & = & F_{0,1}^{(\rm ground)}\cdot {\widehat F}_{0,1}(\tau, \overline{\tau})
\equiv
\left|q^{-k^*/2}\right|^2\cdot
\left|\prod_{n=2}^\infty\frac{1+q^{n-1/2}}
{1-q^n}\right|^2\,, 
\label{F01}
\\
G_{0,1}(\tau, \overline{\tau}) & = & G_{0,1}^{(\rm ground)}\cdot 
{\widehat G}_{0,1}(\tau, \overline{\tau})
\equiv
\left|q^{-k^*/2}\right|^2\cdot
\left|\prod_{n=2}^\infty\frac {1-q^{n-1/2}}{1-q^n}\right|^2\,.
\label{G01}
\end{eqnarray}  
Here the contributions $F_{0,1}^{(\rm ground)}= G_{0,1}^{(\rm ground)}\equiv \left|q^{-k^*/2}\right|^2$ are related to the ground state energy; the contribution $G_{0,1}$ of $X_{0,1}$ is obtained by reversing the sign of all fermionic contributions in (\ref{F01}). 
The complete functions $F(\tau)$, $G(\tau)$ can be computed by summing $F_{0,1}$, $G_{0,1}$ over modular images with $(c+d)$ odd. It corresponds to the spin structure with $\mu=\nu = 1/2$ and $\mu=0$ and $\nu=1/2$, respectively:
\begin{eqnarray}
{\widehat F} {\left[ \begin{array}{c} \frac{1}{2} \\ \frac{1}{2} \end{array} \right]}
(\tau, \overline{\tau}) & = & \sum_{c,d|(c+d)\,\,
{\rm odd}}{\widehat F}_{0,1}((a\tau+b)/(c\tau+d), \overline{\tau})\,,
\label{F11}
\\
{\widehat G} {\left[ \begin{array}{c} 0 \\ \frac{1}{2} \end{array}
\right]    }
(\tau, \overline{\tau}) & = & \,\,\,\,\,\, \sum_{c,d|d\,\,{\rm odd}}
\,\, {\widehat G}_{0,1}((a\tau+b)/(c\tau+d), \overline{\tau})\,.
\label{G11}
\end{eqnarray}
Note that the summand in (\ref{F11}) and (\ref{G11}) does not depend on the choice $a, b$. A modular transformation
$\tau\to\tau+1$ exchanges the pair $(\mu,\nu)=(0,1/2)$ with
$(\mu,\nu)=(1/2,1/2)$; in particular,
$
F(\tau)=G(\tau+1)=F(\tau+2)\,.
$
One can compute the Ramond partition function $K={\rm Tr}_{\rm R}\exp(-\beta H-i\theta J)$ for $\mu=1/2$, $\nu=0$, so
$
K(\tau)=G(-1/\tau).
$ 
This completes the list of three of the four partition functions. In a supersymmetric theory with discrete spectrum, the fourth partition
function $Q=\Tr_{\rm R}(-1)^F\exp(-\beta H-i\theta J)$ is an
integer, independent of $\beta$ and $\theta$ (it can be
interpreted as the index of a supersymmetry generator). 
This function has to be computed using the odd spin structure, the one with $\mu=\nu=0$. 
Typically in three-dimensional gravity the partition function $Q$ vanishes, since the odd spin structure does not extend over any three-manifold with boundary $\Sigma$.

We shall now analyze the partiton function in more detail.
Our goal here will be to repeat the analysis of the previous section by using the spectral functions representation for holomorphycally $\{\cR(s), \cR(\sigma)\}$ and antiholomorphycally $\{\cR(\overline{s}), \cR(\overline{\sigma})\}$
factorized theory. We get
\begin{eqnarray}
\!\!\!\!\!\!\!\!{\widehat F}_{0,1}(\tau, \overline{\tau}) 
& = & 
\left[\frac{\prod_{n=2}^\infty (1+q^{n-1/2})} 
{\prod_{n=2}^\infty (1-q^{n})}\right]
\cdot
\left[\frac{\prod_{n=2}^\infty (1+\overline{q}^{n-1/2})}
{\prod_{n=2}^\infty (1-\overline{q}^{n})} \right]
\nonumber 
\\
\nonumber 
\\
& = &
\left[\frac{\cR(\sigma = 3/2- (3/2)it + i\eta(\tau))}{\cR(s= 2-2it)}\right]\cdot
\left[\frac{\cR(\overline{\sigma}= 3/2+ (3/2)it + i\eta(\tau))}{\cR(\overline{s}=2+2it)}\right]\,,
\label{F}
\\
\nonumber
\\
\nonumber
\\
\!\!\!\!\!\!\!\!\!\!
{\widehat G}_{0,1}(\tau, \overline{\tau}) 
& = & 
\left[\frac{\prod_{n=2}^\infty (1-q^{n-1/2})} 
{\prod_{n=2}^\infty (1-q^{n})}\right]
\cdot
\left[\frac{\prod_{n=2}^\infty (1-\overline{q}^{n-1/2})}
{\prod_{n=2}^\infty (1-\overline{q}^{n})} \right]
\nonumber 
\\
\nonumber
\\
& = &
\left[\frac{\cR(s= 3/2-(3/2)it)}{\cR(s=2-2it)}
\right]\cdot
\left[\frac{\cR(\overline{s}=3/2 + (3/2)it)}{\cR(\overline{s}=2+2it)}
\right]. 
\label{G}
\end{eqnarray}
The complete functions $F(\tau), G(\tau)$ becomes
\begin{eqnarray}
F{\left[ \begin{array}{c} \frac{1}{2} \\ \frac{1}{2} \end{array} \right]}
(\tau, \overline{\tau}) & = & 
\!\!\!\! \sum_{c,d|(c+d)\,\,
{\rm odd}} F_{0,1}^{(\rm ground)}(\gamma\cdot \tau, \overline{\tau})
\,{\widehat F}_{0,1}(\gamma \cdot\tau, \overline{\tau})
\nonumber 
\\
\nonumber
\\
&= & \!\!\!\! \sum_{c,d|(c+d)\,\,
{\rm odd}}
\!\!
F_{0,1}^{(\rm ground)}(\gamma\cdot\tau, \overline{\tau}) 
\left\{\left[\frac{\cR(\sigma = 3/2 - (3/2)it + (1+2d/c)i\eta(\tau))}{\cR(s= 2-2it + i\eta(\tau))}
\right]\right\}_{\gamma} 
\nonumber 
\\
\nonumber
\\
&\times &
\left\{\left[\frac{\cR(\overline{\sigma}= 3/2+ (3/2)it + (1+2d/c)i\eta(\tau))}{\cR(\overline{s}=2+2it + i\eta(\tau))}
\right]\right\}_{\gamma},
\label{F1}
\\
\nonumber
\\
\nonumber
\\
G {\left[ \begin{array}{c} 0 \\ \frac{1}{2} \end{array} \right]}
(\tau, \overline{\tau}) & = & 
\!\!\!\! \sum_{c,d|(c+d)\,\,
{\rm odd}} G_{0,1}^{(\rm ground)}(\gamma\cdot \tau, \overline{\tau})
\,{\widehat G}_{0,1}(\gamma \cdot\tau, \overline{\tau})
\nonumber 
\\
\nonumber
\\
&= &  
\!\!\!\! \sum_{c,d|(c+d)\,\,
{\rm odd}}
\!\! G_{0,1}^{(\rm ground)}(\gamma\cdot \tau, \overline{\tau}) 
\left\{\left[\frac{\cR(s= 3/2 - (3/2)it + (1+2d/c)i\eta(\tau))}
{\cR(s= 2-2it + (2d/c)i\eta(\tau))}
\right]\right\}_{\gamma}
\nonumber 
\\
\nonumber
\\
&\times &
\left\{\left[\frac{\cR(\overline{s}=3/2 + (3/2)it + (1+2d/c)i\eta(\tau))}{\cR(\overline{s}=2+2it + (2d/c)i\eta(\tau))}
\right]\right\}_{\gamma}\,.
\label{G1}
\end{eqnarray}
\begin{remark}
In the case the final sums {\rm (\ref{summand}), (\ref{F1}), (\ref{G1})} are divergent (these kind of divergences have also been encountered in similar sums in 
{\rm \cite{Dijkgraaf3,Kleban,Manschot})}
the one-loop corrections has to be regularized. This procedure can be developed by the Poisson summation technique in a way similar to the calculations of Poincar\'{e} series in {\rm \cite{Maloney}}, Sect. {\rm 3.2} {\rm (}see also {\rm \cite{Iwaniec}}{\rm )}. It is a natural regularization, a type of the zeta function regularization. We hope we will return to this interesting problem in a forthcoming paper. In this and next sections we leave this question open.
\end{remark}

\section{Homological aspects of combinatorial identities}
\label{Outline}

Before considering other examples of partition functions we would like now to spend some
time on what we believe is the mathematical origin of the combinatorial identities that are
at the basis of the relation between partition function and formal power series and homologies of Lie algebras.  
The following is meant to be a brief introduction to homological aspects of differential complexes,
in particular we would like to show how combinatorial identities could be derived from initial complex of (graded) Lie algebras. In recalling the main results we shall follow the book \cite{Fuks}. Our interest is the Euler-Poincar{\'e} formula associated with a complex consisting of finite-dimensional linear spaces. The relationship between Lie algebras and combinatorial identities was first discovered by Macdonald \cite{Macdonald,Macdonald1}; the Euler-Poincar{\'e} formula
is useful for combinatorial identities known as {\it Macdonald identities}. The Macdonald identities are related to Lie algebras in one way or another and can be associated with generating functions (in particular, elliptic genera) in quantum theory. 
 
Let ${\mathfrak g}$ be an {\it infinite-dimensional} Lie algebra, and assume that this Lie algebra possesses a grading, i.e. ${\mathfrak g}$ is a direct sum of its homogeneous component 
${\mathfrak g}_{(\lambda)}$, where $\lambda$ are elements of an abelian group; $[{\mathfrak g}_{(\lambda)}, {\mathfrak g}_{(\mu)}]
\subset {\mathfrak g}_{(\lambda + \mu)}$ (recall for example the Virasoro algebra). Let us consider  a module $A$ over ${\mathfrak g}$, or ${\mathfrak g}$-module. $A$ is a vector with the property that that there  exists a bilinear map $\mu : {\mathfrak g}\times A \rightarrow A$ such that
$[{\mathfrak g}_1, {\mathfrak g}_2] ={\mathfrak g}_1({\mathfrak g}_2a) - {\mathfrak g}_2 ({\mathfrak g}_1a)$ for all $a\in A,\, {\mathfrak g}_1, {\mathfrak g}_2 \in {\mathfrak g}$ occurs.
In other words our ${\mathfrak g}$-module is a left module over the
universal enveloping algebra $U({\mathfrak g})$ of ${\mathfrak g}$.
\footnote{
$U({\mathfrak g})$ is the quotient algebra of the tensor
algebra $T^n{\mathfrak g} =
\bigoplus_{n=0}^{\infty}
(\overbrace{{\mathfrak g}\otimes ...
\otimes {\mathfrak g}}^{n-\mbox{times}})$
by the ideal generated by
elements of the form
$[{\mathfrak g}_1, {\mathfrak g}_2] - ({\mathfrak g}_1\otimes{\mathfrak g}_2 -
{\mathfrak g}_2 \otimes {\mathfrak g}_1)$.
}
Let $C^n({\mathfrak g}; A)$ be the space of all cochains, an
$n$-dimensional cochain of the algebra ${\mathfrak g}$ with coefficients in $A$ being a skew-symmetric $n$-linear functional on ${\mathfrak g}$ with values in $A$. 
Since $C^n({\mathfrak g}; A) = {\rm Hom} (\Lambda^n, A)$,
the cochain space $C^n({\mathfrak g}; A)$ becomes a
${\mathfrak g}$-module.
The differential $d= d_n : C^n({\mathfrak g}; A)\rightarrow
C^{n+1}({\mathfrak g}; A)$ can be defined as follows
\begin{eqnarray}
dc({\mathfrak g}_1, \ldots ,{\mathfrak g}_{n+1})
& = &
\sum_{1\leq s\leq t\leq n+1}(-1)^{s+t-1}c
([{\mathfrak g}_s, {\mathfrak g}_t], {\mathfrak g}_1, \ldots , {\widehat {\mathfrak g}}_s, \ldots
{\widehat {\mathfrak g}}_t, \ldots,{\mathfrak g}_{n+1})
\nonumber \\
& + &
\sum_{1\leq s\leq n+1}(-1)^{s}{\mathfrak g}_sc
({\mathfrak g}_1, \ldots,{\widehat{\mathfrak g}}_s,...,{\mathfrak g}_{n+1})
\mbox{,}
\label{diff}
\end{eqnarray}
where $c\in C^n({\mathfrak g}; A),\, {\mathfrak g}_1, \ldots, {\mathfrak g}_{n+1}
\in {\mathfrak g}$, and $C^n({\mathfrak g}; A) =0, \, d_n= 0$ for $n<0$.
Since $d_{n+1}\circ d_n =0$ for all $n$, the set
$C^{\bullet}({\mathfrak g}; A) \equiv \{C^n({\mathfrak g}; A), d_n\}$ is an algebraic complex, while the corresponding cohomology
$H^n({\mathfrak g}; A)$ is referred to as the cohomology of the
algebra ${\mathfrak g}$ with coefficients in $A$.
Let $C_n({\mathfrak g}; A)$ be the space of $n$-dimensional chains of the Lie algebra ${\mathfrak g}$. It can be defined as
$A\otimes \Lambda^n{\mathfrak g}$. The differential $\delta= \delta_n:
C_n({\mathfrak g}; A)\rightarrow C_{n-1}({\mathfrak g}; A)$ is defined by the formula
\begin{eqnarray}
\delta (a\otimes ({\mathfrak g}_1 \wedge \ldots \wedge {\mathfrak g}_{n}))
& = &
\sum_{1\leq s\leq t\leq n}(-1)^{s+t-1}a\otimes
([{\mathfrak g}_s, {\mathfrak g}_t]\wedge {\mathfrak g}_1\wedge \ldots
{\widehat {\mathfrak g}}_s \ldots {\widehat {\mathfrak g}}_t \ldots \wedge
{\mathfrak g}_{n})
\nonumber \\
& + &
\sum_{1\leq s\leq n}(-1)^{s}{\mathfrak g}_sa\otimes
({\mathfrak g}_1\wedge \ldots {\widehat{\mathfrak g}}_s \ldots \wedge 
{\mathfrak g}_{n})
\mbox{.}
\label{diff1}
\end{eqnarray}
The homology $H_n({\mathfrak g}; A)$ of the complex $\{C_n({\mathfrak g}; A), \delta_n\}$ is referred to as the homology of the algebra ${\mathfrak g}$. 
Suppose now that the $\mathfrak g$-module $A$ can be graded by homogeneous components $A_{(\mu)}$ in such way that ${\mathfrak g}_{(\lambda)}A_{(\mu)}\subset A_{(\lambda+\mu)}$. In the case when the module $A$ is trivial we assume that $A=A_{(0)}$.
The grading of our Lie algebra endows with a grading both the chain and cochain spaces, for we can define
\begin{eqnarray}
C^n_{(\lambda)}({\mathfrak g}; A) & = & 
\{c\in C^n({\mathfrak g}; A)|
c({\mathfrak g}_1, \ldots,{\mathfrak g}_n)\in
A_{(\lambda_1+ \ldots +\lambda_n-\lambda)} \,\,\,\,\,
{\rm for}\,\,\,\,\,{\mathfrak g}_i\in {\mathfrak g}_{(\lambda_i)}\};
\nonumber \\
C_n^{(\lambda)}({\mathfrak g}; A) & {\rm is} & \, {\rm
generated}\,\,\, {\rm by}\,\,\, {\rm the}\,\,\, {\rm
chains}\,\,\, a\otimes({\mathfrak g}_1\wedge \ldots \wedge
{\mathfrak g}_n),\, a \in A_{(\mu)},
\nonumber \\
& {\mathfrak g}_i & \!\!\!\!\in {\mathfrak g}_{(\lambda_i)}, \,\,\,
\lambda_1+ \ldots +\lambda_n+\mu = \lambda \,. 
\label{chain}
\end{eqnarray}
We get $d(C^n_{(\lambda)}({\mathfrak g}; A)) \subset
C^{n+1}_{(\lambda)}({\mathfrak g}; A)$ and
$\delta (C_n^{(\lambda)}({\mathfrak g}; A)) \subset
C_{n-1}^{(\lambda)}({\mathfrak g}; A)$ and both spaces acquire gradings\footnote{
The cohomological (and homological) multiplicative structures
are compatible with these gradings, for example,
$H^m_{(\lambda)}({\mathfrak g})\otimes H^n_{(\mu)}({\mathfrak g})\subset
H^{m+n}_{(\lambda + \mu)}({\mathfrak g})$.}.
The chain complex
$C_{\bullet}({\mathfrak g})$,\,
${\mathfrak g}= \oplus_{\lambda=1}^{\infty}{\mathfrak g}_{(\lambda)},\,
{\rm dim}\,{\mathfrak g}_{(\lambda)}< \infty$,
can be decomposed as follows:
\begin{equation}
0\longleftarrow C_{0}^{(\lambda)}({\mathfrak g}) \longleftarrow
C_{1}^{(\lambda)}({\mathfrak g}) \ldots\longleftarrow C_N^{(\lambda)}({\mathfrak g}) \longleftarrow 0\,.
\end{equation}
The Euler-Poincar{\'e} formula gives:
\begin{equation}
\sum_m(-1)^m {\rm dim}\,C_m^{(\lambda)}({\mathfrak g}) = 
\sum_m(-1)^m {\rm dim}\,H_m^{(\lambda)}({\mathfrak g})\,.
\label{E-P}
\end{equation}
{ As a consequence, we can
introduce the $q$ variable and rewrite identity (\ref{E-P}) as a formal power series}
\footnote{
If $A$ is main field the notation $C_n({\mathfrak g}; A),\, H_n({\mathfrak g}; A)$ is abbreviated to $C_n({\mathfrak g}),\, H_n({\mathfrak g})$.
For a finite-dimensional algebra $\mathfrak g$ we obviously have $C^n({\mathfrak g})=[C_n({\mathfrak g})]^{*}$ and $H^n({\mathfrak g})=[H_n({\mathfrak g})]^{*}$, where a simbol $*$ denotes the dual space. It is clear that in the case of cohomologies we have a power series similar to (\ref{EP}). For the finite-dimensional case, as well as the infinite-dimensional case, an element of the space $H^n({\mathfrak g}; A)$ defines a linear map $H_n({\mathfrak g})\rightarrow A$; if the algebra $\mathfrak g$ and the module $A$ are finite dimensional, then $H^n({\mathfrak g}; A^{*})= [H_n({\mathfrak g}; A)]^{*}$.
}:
\begin{equation}
\sum_{m,\lambda}(-1)^m q^\lambda {\rm dim}\,C_m^{(\lambda)}({\mathfrak g}) = \sum_{m,\lambda}(-1)^m q^\lambda {\rm dim}\,H_m^{(\lambda)}({\mathfrak g})
=\prod_n(1-q^n)^{{\rm dim}\,{\mathfrak g}_{n}}\,.
\label{EP}
\end{equation}
In order to get the identity in its final form the
homology $H_m^{(\lambda)}({\mathfrak g})$ has to be computed. 
Let ${\mathfrak g}$ be the (poly)graded Lie algebra,
$
{\mathfrak g} = \bigoplus_{\scriptstyle \lambda_1\geq 0, ..., \lambda_k\geq 0
\atop\scriptstyle \lambda_1+...+\lambda_k>0}
{\mathfrak g}_{(\lambda_1, ..., \lambda_k)},
$ 
satisfying the condition
$
{\rm dim}\, {\mathfrak g}_{(\lambda_1, ..., \lambda_k)} < \infty. 
$ 
For formal power series in $q_1,...,q_k$ we have the following identity \cite{Fuks}:
\begin{equation}
\sum_{m,\lambda_1,...,\lambda_k}(-1)^m 
q_1^{\lambda_1}...q_k^{\lambda_k} H_m^{(\lambda_1,...,\lambda_k)}
=
\prod_{n_1,...,n_k}\left(1-q_1^{n_1}\cdots q_k^{n_k}\right)^
{{\rm dim}\, {\mathfrak g}_{n_1, ..., n_k}}\,.
\end{equation}
The reader can find in 
\cite{Fuks} some computations of the homology $H_m^{(\lambda)}$ in its final form . There are algebras which can be constructed from exterior authomorphisms of simple algebras possessing a finite center. In this case one must find the dimensions of the space $H_\bullet^{(\lambda_1,...,\lambda_n)}(\mathfrak g)$. This is not too difficult to do, but the answer turns out to be rather cumbersome.

We also consider free associative Lie algebras. The importance of these algebras follows from concepts of deformation theory. Let $\bf k$ denote a commutative and associative ring with a unit (all modules and algebras are taken over $\bf k$).
Define a magma ${\mathfrak M}$ as a set with a map ${\mathfrak M}\times
{\mathfrak M} \rightarrow {\mathfrak M}$ (denoted by $(x,y)\mapsto xy$).
For a set $X$ we also define a
family of sets $X_n,\, n \geq 1$ as follows: $X_1=X,\, X_n =
\coprod_{\ell+m=n}(X_\ell\times X_m)$\, (disjoint union),\, $n\geq 2$. Put ${\mathfrak M}_X = \coprod_{n=1}^{\infty} X_n$ and
let us define ${\mathfrak M}_X\times
{\mathfrak M}_X \rightarrow {\mathfrak M}$ by means of 
\begin{equation}
X_\ell\times X_m
\stackrel{{\rm inclusion}}{\longrightarrow} X_{\ell+m}\subset {\mathfrak M}_X \,\,\,\,\, ({\rm the}\,\,\,{\rm magma}\,\,\, {\mathfrak M}_X\,\,\,{\rm is}\,\,\,{\rm called }\,\,\,
{the}\,\,\, {free}\,\,\,
{magma}\,\,\,{\rm on}\,\,\, X)\,.
\end{equation}
Let ${\mathfrak A}_X$ be the $\bf k$-algebra of the free magma ${\mathfrak M}_X$.
An element $\alpha \in {\mathfrak A}$ is a finite sum 
$\alpha = \sum_{n\in {\mathfrak M}_X}nC_n,\, C_n\in {\bf k}$, and the multiplication in ${\mathfrak A}_X$ extends the multiplication in ${\mathfrak M}_X$. The $\bf k$-algebra ${\mathfrak A}_X$ of the free magma ${\mathfrak M}_X$ is called the free algebra on $X$. 
Let ${\mathfrak I}_X$ be the two-side ideal of ${\mathfrak A}_X$, the quotient algebra ${\mathfrak g}_X\equiv {\mathfrak A}_X/{\mathfrak I}_X$ is called the free Lie algebra on $X$. ${\mathfrak I}_X$ is a graded ideal of ${\mathfrak A}_X$, which means that ${\mathfrak g}_X$ has a natural structure of graded algebra.
Let ${\mathcal V} = {\bf k}^{(X)}$ be the free $\bf k$-module with basis $X$.
The free associative algebra on $X$, denoted by $\cA_{X}$, is the tensor algebra $T^n {\mathcal V}$ of $\mathcal V$
\footnote{
Recall that for any Lie algebra ${\mathfrak g}$ over commutative ring $\bf k$ and the tensor algebra 
$T^n{\mathfrak g}$, one has 
${\rm Hom}_{\rm mod}({\mathfrak g}{\mathfrak A}) = 
{\rm Hom}_{\cA_X}(T^n{\mathfrak g}{\mathfrak A})$.
}.
Assume $\mu : {\mathfrak g}_X\rightarrow \cA_X$ to be the map induced by the map $X\rightarrow \cA_X$, then the map $\mu$ is an
isomorphism of ${\mathfrak g}_X$ onto the Lie subalgebra of $\cA_X$
generated by $X$. ${\mathfrak g}_X$ and its
homogeneous components ${\mathfrak g}_X^n$ are free 
$\bf k$-module. If $X$ is finite and ${\mathfrak g}_X^n$ is free of finite rank ${\ell}_{({\rm card}\, X)}(n)$ 
\footnote{
Let $\omega \in {\mathfrak M}_X$, then $\omega$ is called a non-associative word on $X$; its length, $\ell (\omega)$, is the unique $n$ such that $\omega \in X_n$.
}
\,\,then
$
\sum_{n|m}n{\ell}_{({\rm card}\,X)}(n) = ({\rm card}\,X)^m 
\mbox{.}
$
Let us assume that $\bf k$ is a field, and choose a homogeneous basis $\{{\mathfrak g}_j\}_{j\in I_X}$ of ${\mathfrak g}_X$. For any positive integer $n$ we have that in the product
$
\prod_{j\in {\mathfrak I}_X}(1-q^{{\rm deg}\,{\mathfrak g}_j })^{-1}
$
the number of factors such that ${\rm deg}\,{\mathfrak g}_j = n$
is the rank ${\ell}_{({\rm card}\,X)}(n)$ of ${\mathfrak g}_X^n$
{\rm \cite{Serre}}, and then 
\begin{equation}
\prod_{j\in {\mathfrak I}_X}(1-q^{{\rm deg}\,{\mathfrak g}_j })^{-1}
= \prod_{n=1}^{\infty}(1-q^{n})^{-{\ell}_{({\rm card}\,X)}(n)}\,.
\end{equation}
Since $\cA_X$ is the free associative Lie algebra on $X$ the
family of monomials $x_{j_1}\ldots x_{j_n}$,\, $x_{j_k}\in X$ is a
basis of $\cA_X^n$, and therefore ${\rm rank}(\cA_X^n) = ({\rm card}\,X)^n$.
\\

\noindent
\begin{remark} In general one can obtain certain formulae for Poincar{\'e} polynomials in the following form
\begin{equation}
\prod_n(1-q^n)^{{\rm dim}\,{\mathfrak g}_n},\,\,\,\,\,\,\,\,\,\,\,\,
\prod_n(1-q^n)^{{\rm rank}\,{\mathfrak g}_n}\,.
\label{General}
\end{equation}
These formulae are associated with dimensions of homologies of appropriate topological spaces and linked to generating functions and, possibly, to elliptic genera. In the next section we will show that generating functions of the form {\rm (\ref{General})} admit representations in term of spectral Selberg-type functions related to the congruence subgroup of $SL(2, {\mathbb Z})$.
\end{remark}

\section{Elliptic genera of generalized symmetric products of $N$-folds}
\label{N-Folds}

In this section we show on examples that particular cases of (\ref{General}) can be written in terms of Selberg-type (Ruelle) spectral functions associated with $q$-series.
We concentrate specifically on nonlinear sigma models. But before going to the main topic of this section we would like to premise a short intruction about Heisenberg algebras and Hilbert schemes.

\subsection{Hilbert schemes and Heisenberg algebras}
\label{HH}

Preliminary to the subject of symmetric products and their connection with spectral functions, we explain briefly the relation between the Heisenberg algebra and its representations, and the Hilbert scheme of points, mostly following the lines of \cite{Nakajima00}. To be more specific:
\begin{itemize}
\item{}
The infinite dimensional Heisenberg algebra (or, simply, the Heisenberg algebra) plays a fundamental role in the representation theory of the affine Lie algebras. An important representation of the Heisenberg algebra is the Fock space representation on the polynomial ring of infinitely many variables. The degrees of polynomials (with different degree variables) give a direct sum decomposition of the representation, which is called weight space decomposition.
\item{}
The Hilbert scheme of points on a complex surface appears in the algebraic geometry. The Hilbert scheme of points decomposes into infinitely many connected components according to the number of points. Betti numbers of the Hilbert scheme have been computed in \cite{Gottsche}. The sum of the Betti numbers of the Hilbert scheme of $N$-points is equal to the dimension of the subspaces of the Fock space representation of degree $N$.
\end{itemize}
\begin{remark}
If we consider the generating function of the Poincar\'{e} polynomials associated with set of points we get the character of the Fock space representation of the Heisenberg algebra. 
The character of the Fock space representation of the Heisenberg algebra {\rm (}in general the integrable highest weight representations of affine Lie algebras{\rm )} are known to have modular invariance as has been proved in {\rm \cite{Kac2}}. This occurrence is naturally explained through the relation to partition functions of conformal field theory on a torus. In this connection
the affine Lie algebra has close relation to conformal field theory. 
\end{remark}

\noindent
{\bf Algebraic preliminaries.}
Let ${R} = {\mathbb Q}[p_1, p_2, ...]$
be the polynomial ring of infinite many variables $\{p_j\}_{j=1}^{\infty}$. Define $P[j]$ as $j\partial/\partial p_j$
and $P[- j]$ as a multiplication of $p_j$ for each positive $j$. Then the commutation relation holds:
$
[\,P[i],\, P[j]\,] = i\delta_{i+j, 0}\,{\rm Id}_{R},\,
$ 
$i, j \in {\mathbb Z}/\{0\}.$ We define the infinite dimensional Heisenberg algebra as a Lie algebra generated by $P[j]$ and $K$ with defining relation
\begin{equation}
[\,P[i],\, P[j]\,] = i\delta_{i+j, 0} K_{R},\,\,\,\,\,
[\,P[i],\, K\,] = 0,\,\,\,\,\, i, j \in {\mathbb Z}/\{0\}.
\end{equation}
The above ${R}$ labels the representation.
If $1\in {R}$ is the constant polynomial, then $P[i]1 = 0,\, i\in {\mathbb Z_+}$ and
\begin{equation}
{R} =  {\rm Span}\{ P[-j_1] \cdots  P[-j_k]\,1\mid
k\in {\mathbb Z}_+\cup \{0\}, \,\,\, j_1, \dots, j_k \in 
{\mathbb Z}_+\}\,.
\end{equation}
1 is a highest weight vector. 
This is known in physics as the {\it bosonic Fock space}. 
The operators $P[j]\, (j< 0)$ ($P[j]\, (j>0)$) are the {\it creation (annihilation) operators}, while 1 is  the {\it vacuum vector}.

Define the degree operator 
${\cD}: {R}\rightarrow {R}$ by \,
$
{\cD}(p_1^{m_1} p_2^{m_2}\cdot\cdot\cdot \,)\stackrel{def}{=}
(\sum_i i m_i) p_1^{m_1} p_2^{m_2}\ldots
$
The representation ${R}$ has $\cD$ eigenspace decomposition; the eigenspace with eigenvalue $N$ has a basis
$
p_1^{m_1} p_2^{m_2}\cdot\cdot\cdot
(\sum_i i m_i) = N.
$
Recall that partitions of $N$ are defined by a nonincreasing sequence of nonnegative integers $\nu_1\geq \nu_2 \geq \ldots$ such that $\sum_\ell \nu_\ell = N$. One can represent $\nu$ as $(1^{m_1}, 2^{m_2}, \cdot\cdot\cdot)$ (where 1 appears $m_1$-times, 2 appears $m_2$-times, \ldots in the sequence). Therefore, elements of the basis corresponds bijectively to a partition $\nu$. The generating function of eigenspace dimensions, or the {\it character}
in the terminology of the representation theory, is well--known to have the form
\begin{equation}
{\rm Tr}_{R}\, q^{\cD}\, \stackrel{def}{=}
\sum_{N\in {\mathbb Z}_+\cup \{0\}} 
q^N {\rm dim}\,\{ r\in {R}\, \mid \, {\cD}r = Nr\,\}
= \prod_{n= 1}^{\infty}(1 - q^n)^{-1}\,.
\end{equation}

Let us define now the Heisenberg algebra associated with a finite dimensional $\mathbb Q$-vector space $V$ with non-degenerate symmetric bilinear form $(\, ,\, )$. Let $W = (V\otimes t\,{\mathbb Q}[t])\oplus (V\otimes t^{-1}\,{\mathbb Q}[t^{-1}])$, then define a skew-symmetric bilinear form on $W$ by
$(r\otimes t^i,\,s\otimes t^j) = i \delta_{i+j, 0} (r, s)$.
The Heisenberg algebra associated with $V$ can be defined as follows: we take the quotient of the free algebra $L(W)$ divided
by the ideal $\mathfrak I$ generated by $[r,\, s] - (r,\,s)1\,\, (r, s\in W)$. It is clear that when $V = {\mathbb Q}$ we have the above Heisenberg algebra. For an orthogonal basis 
$\{ r_j\}_{j=1}^p$ the Heisenberg algebra associated with $V$ is isomorphic to the tensor product of $p$-copies of the above Heisenberg algebra.

Let us consider next the {\it super}-version of the Heisenberg algebra, the super-Heisenberg algebra. The initial data are a vector space 
$V$ with a decomposition $V = V_{\rm even}\oplus V_{\rm odd}$
and a non-degenerate bilinear form satisfying 
$(r, s) = (-1)^{|r||s|}(r, s)$. In this formula $r, s$ are either elements of $V_{\rm even}$ or $V_{\rm odd}$, while $|r| = 0$ if
$r \in V_{\rm even}$ and $|r|=1$ if $r\in V_{\rm odd}$. 
As above we can define $W$, the bilinear form on $W$, and $L(W)/{\mathfrak I}$, where now we replace the Lie bracket 
$[\, ,\,]$ by the super-Lie bracket. In addition, to construct the free-super Lie algebra in the tensor algebra we set
$
(r\otimes t^i,\,s\otimes t^j) = (r \otimes t^i)(s \otimes t^j)
+ (s \otimes t^j)(r \otimes t^i)
$
for $r, s \in V_{\rm odd}$.
By generalizing the representation on the space of polynomials of infinite many variables one can get a representation of the super-Heisenbwerg algebra on the symmetric algebra ${R}= S^*(V \otimes t\, {\mathbb Q} [t])$ of the positive degree part $V\otimes t\, {\mathbb Q}\,[t]$. As above we can define the degree operator $\cD$. The following character formula holds:
\begin{equation}
{\rm Tr}_{R}\, q^{\cD} = \prod_{n = 1}^{\infty}
\frac{(1 + q^n)^{{\rm dim}\, V_{\rm odd}}}
{(1 - q^n)^{{\rm dim}\, V_{\rm even}}}
= 
\frac{\cR(\sigma = 1-it+ i\eta(\tau))^{{\rm dim}\, V_{\rm odd}}}
{\cR(s = 1-it)^{{\rm dim}\, V_{\rm even}}}\,. 
\label{trace-usual}
\end{equation}
Counting the odd degree part by $-1$ we can replace the usual trace by the super-trace (denoted by ${\rm STr}$) and get the result
\footnote{
In the case when $V$ has one-dimensional odd degree part only
(the bilinear form is $(r, r) =1$ for a nonzero vector $r\in V$) the above condition is not satisfied. We can modify the definition of the corresponding super-Heisenberg algebra by changing the bilinear form on $W$ as $(r\otimes t^i, r\otimes t^j) = \delta_{i+j, 0}$. The resulting algebra is called {\it infinite dimensional Clifford algebra}. The above representation $R$ can be modified as follows and it is the {\it fermionic Fock space} in physics. The representation of the even degree part was realized as the space of polynomials of infinity many variables; the Clifford algebra is realized on the exterior algebra 
${R} = \wedge^*(\bigoplus_j{\mathbb Q} dp_j)$ of a vector space with a basis of infinity many vectors. For $j>0$ we define $r\otimes t^{- j}$ as an exterior product of $dp_j$, $r\otimes t^j$ as an interior product of $\partial/\partial p_j$.
} 
\begin{equation}
{\rm STr}_{R}\, q^{\cD} = \prod_{n = 1}^{\infty}
(1 - q^n)^{{\rm dim}\, V_{\rm odd} - {\rm dim}\, V_{\rm even}}
= \cR(s = 1-it)^{{\rm dim}\, V_{\rm odd} - {\rm dim}\, V_{\rm even}}\,. 
\end{equation}
\\

\noindent
{\bf Geometric interpretation.} 
Suppose that $X$ is a nonsingular quasi-projective surface defined over the complex number field $\mathbb C$. 
Let us consider the configuration space of $N$ distinct unordered points $x_1, \ldots , x_N$. Suppose that with these points we can parametrize geometric objects where some points are allowed to collide. Naively we can consider just the structure as subsets. As a result we get the $N$th symmetric product $\CS^NX$ (see for notation Sect. \ref{Sigma}) where the symmetric group $\CS_N$ acts on $X^N$ as a permutation of entries. On the other hand for the Hilbert scheme we do not consider collections of points merely as subsets, but we also consider functions on those.
Thus let $X^{[N]}$ be the Hilbert scheme parametrizing $0$-dimensional subschemes of $X$ with length $N$. We recall that $X^{[N]}$ is the set of ideal sheaves ${\mathfrak J}\subset \cO_X$ of the structure sheaf $\cO_X$, whose Hilbert polynomials are constant functions $N$. Define a holomorphic map $\pi: X^{[N]}\rightarrow \CS^NX$. The so-called Hilbert-Chow morphism, from the Hilbert scheme to the symmetric products, is defined as follows. For a given ideal ${\mathfrak J} \in X^{[N]}$,
consider the quotient sheaf $\cO_X/{\mathfrak J}$, which has support in a finite set of points. Suppose that the length $(\cO_X/{\mathfrak J})_x$ is the lenght
of $\cO_X/{\mathfrak J}$ at the point $x$ in the support. The Hilbert-Chow morphism is:
\begin{equation}
\pi (\mathfrak J) \stackrel{def}{=}
\sum_{x \in {\rm Supp}\,(\cO_X/{\mathfrak J})} {\rm length}\,
(\cO_X/{\mathfrak J})_x [x]\,.
\label{pi}
\end{equation}
In the right hand side of (\ref{pi}) a point in the symmetric  product is represented in the summation notation $\sum_i[x_i]$.
\\

\noindent
{\bf Example}: Let $N=2$. If $x_1, x_2$ are different points in $X$,
the sheaf of holomorphic functions on $\cF =\{x_1, x_2\}$ is the quotient of the structure sheaf $\cO_X$ divided by the sheaf of functions vanishing at $x_1$ and $x_2$. The set of functions vanishing at $x_1$ and $x_2$, when $x_2$ approhes $x_1$, converges to the set of functions that vanish at $x_1$
together with their derivatives in the direction in which $x_2$ approaches $x_1$: 
\begin{equation}
\{f \in \cO_X \,\mid \, f(x_1) = 0,\,\,\,\, df_{x_1}(r) =0 \}\,.
\label{f}
\end{equation}
Thus $X^{[N]}$ contains two type of elements: the first type is $\{x_1, x_2\}$ and $x_1 \neq x_2$, and the second type is a pair of $x\in X$ and a one-dimensional subspace $\Sigma$ of $T_xX$.
Moreover $\Sigma$ is the subspace spanned by $r$, and this $\Sigma$ makes the difference of the symmetric product $\CS^NX$
and the Hilbert scheme. We have $\pi (\cF) = [x_1] + [x_2]$ for $\cF = \{x_1, x_2\}$ and $\pi (\cF) = 2[x_1]$ for $\cF$ of the form (\ref{f}). 
\\

\noindent
{\bf One-dimensional higher variety.}
For a one-dimensional higher variety (i.e. for a surface) the following results hold:
\begin{itemize}
\item{}
For a Riemann surface (${\rm dim}\,X = 1$) $\CS^NX$ and $X^{N}$
are isomorphic under Hilbert-Chow morphism. 
\item{} 
If $X$ is a nonsingular quasi-projective surface, the Hilbert-Chow morphism $\pi: \, X^{[N]}\rightarrow \CS^NX$ gives a resolution of singularities of the symmetric product $\CS^NX$ \cite{Fogarty}. In particular $X^{[N]}$ is a nonsingular quasi-projective variety of dimension $2N$.
\item{}
If $X$ has a symplectic form, $X^{[N]}$ also has a symplectic form.
For $N= 2$ it has been prooved in \cite{Fujiki}, for $N$ general in \cite{Beauville}.
\item{}
The generating function of the Poincar\'{e} polynomials $P_t(X^{[N]})$ of $X^{[N]}$ is given by
\begin{eqnarray}
\!\!\!\!\!\!\!\!\!\!\!
\sum_{N=0}^{\infty}q^N\, P_t(X^{[N]})
& = &
\prod_{n=1}^{\infty} \frac{(1+t^{2n-1}q^n)^{b_1(X)}(1+t^{2n+1}q^n)^{b_3(X)}}
{(1-t^{2n-2}q^n)^{b_0(X)}(1-t^{2n}q^n)^{b_2(X)}
(1-t^{2n+2}q^n)^{b_4(X)}}
\nonumber \\
& = &
\frac{\prod_{j=0, 2}\,\,\,\,\cR(\sigma = (1+(j-1){\rm log}\,t)(1-it)+ 
i\eta (\widetilde{\tau}))^{b_{j+1}(X)}}
{\!\!\!\! \!\!\!\!\prod_{j=0, 2, 4}\cR(s = (1+(j-2){\rm log}\,t)(1-it)(1-it))^{b_j(X)}}\,.
\label{GFP}
\end{eqnarray}
In the last line of Eq. (\ref{GFP}) we put  
${\rm Im}\,\widetilde{\tau} = {\rm Im}\,\tau - (2\pi)^{-1}
{\rm log}\,t$.
\end{itemize}

\subsection{The elliptic genus of a supersymmetric sigma model on symmetric products}
\label{Sigma}

As was shown in \cite{Dijkgraaf1}, the elliptic genus of an ${\mathcal N}=2$ supersymmetric sigma model on the $N$-fold symmetric product ${\mathfrak S}^NX$ equates to the partition function of a second quantized string theory on the space with $S^1$-action
\footnote{
The reader can find a string theory interpretation of a supersymmetric sigma model on ${\mathfrak S}^NX$ in Sect. 1.1 of \cite{Dijkgraaf1}.
}.
In string compactifications on manifolds of the form $X\times S^1$, one can consider the configuration of a $D$-string wound $N$ times around $S^1$, bound to a $({\rm dim}\,X+1)$-brane. The quantum mechanical degrees of freedom of this $D$-brane configuration are naturally encoded in a two-dimensional sigma model on the $N$-fold symmetric tensor product of $X$, which describes the transversal fluctuations of the $D$-string. This construction is originally due to \cite{Vafa95, Bershadsky95,Strominger1}.

With this physical motivation let us consider a sigma model on the $N$-fold
symmetric product ${\mathfrak S}^NX$ of a K\"{a}hler manifold $X$, that is the $\CS^N X = [X^N/\CS_N] := \underbrace{X\times\cdots\times X}_N /{\mathfrak S}_N$ orbifold space,
$\CS_N$ being the symmetric group of $N$ elements. 
Objects of the category of the {\rm orbispace} $[X^N/{\mathfrak S}_N]$ are  the $N$-tuples $(x_1,\ldots,x_N)$ of points in $X$;  arrows are elements of the form $(x_1,\ldots,x_N; \sigma)$ where $\sigma \in \CS_N$. In addition the arrow $(x_1,\ldots,x_N; \sigma)$ has as its source $(x_1,\ldots,x_N)$, and as its target $(x_{\sigma(1)},\ldots,x_{\sigma(N)})$. 
This category is a groupoid for the inverse of $(x_1,\ldots,x_N; \sigma)$ is $(x_{\sigma(1)},\ldots,x_{\sigma(N)}; \sigma^{-1})$. 
(Orbispace as a groupoid has been describes in \cite{Kontsevich,Moerdijk}).

For a K\"{a}hler manifold $X$ and the orbifold space 
${\mathfrak S}^NX$ the genus one partition function depends on the boundary conditions imposed on the fermionic fields. For definiteness, following \cite{Dijkgraaf1}, we choose the boundary conditions such that the partition function 
$\chi({\mathfrak S}^NX; q, y)$ coincides with the elliptic genus, which is defined as the trace over the Ramond-Ramond (RR) sector of the sigma model of the evolution operator $q^{H}$ times $(-1)^Fy^{F_L}$. In addition $F=F_L + F_R$ is the sum of the left- and right-moving fermion number. If $\cE$ is an elliptic curve with modulus $\tau$ and a line bundle labeled by 
$z\in Jac (\cE)\cong \cE$, then we define $q := e^{2\pi i\tau}$,
$y := e^{2\pi i z}$. The elliptic genus is defined as \cite{Dijkgraaf1}:
\begin{equation}
\chi(X; q, y) = {\rm Tr}_{\cH(X)}(-1)^Fy^{F_L}
q^{L_0-d/8}\overline{q}^{\overline{L}_0-d/8}\,, 
\label{char}
\end{equation}
where $d$ is the complex dimension of a K\"{a}hler manifold $X$ and $\cH(X)$ is the Hilbert space of the ${\cal N}=2$ supersymmetric field theory with target space $X$. 
Note that the space $\oplus_NH^\bullet(X^{[N]})$ forms a representation of the Heisenberg algebra generated by sigma model creation operators $\alpha_n^I$, where $I$ runs over a basis of $H^\bullet(X)$.
Formula (\ref{char}) has a natural physical interpretation. Indeed, it has the form of a trace over an irreducible representation of the Virasoro algebra. The representation contains a ground state $|0\rangle$ of weight $k$,  $L_{0}|0\rangle = -k|0\rangle$, along with its Virasoro descendants $L_{-n_{1}}\dots L_{-n_{i}} |0\rangle$.
\\

\noindent
{\bf Generating functions.}
The Hilbert space of an orbifold field theory can be decomposed into twisted
sectors ${\cH}_{\gamma}$, which are labelled by the
conjugacy classes
$\{\gamma\} $ of the orbifold group $\CS_N$
\cite{Dixon,Dijkgraaf1}.
For a given twisted sector one keeps the states that are invariant under the
centralizer subgroup $\Gamma_{\gamma}$ of the element $\gamma$.
Let ${\cH}_{\gamma}^{\Gamma_{\gamma}}$ be an invariant subspace
associated with $\Gamma_{\gamma}$. One can compute the conjugacy classes $[\gamma]$ by using a set of partitions $\{N_n\}$ of $N$, namely $\sum_nnN_n=N$, where $N_n$ is the multiplicity of
the cyclic permutation $(n)$ of $n$ elements in
the decomposition of $\gamma$: 
$
[\gamma]=\prod_{n=1}^{s}(n)^{N_n}.
$
For this conjugacy class the centralizer subgroup of a permutation
$\gamma$ is \cite{Dijkgraaf1}
\begin{equation}
\Gamma_{\gamma}= \CS_{N_1}\times \prod_{j=2}^s
(\CS_{N_j}\rtimes {\mathbb Z}_j^{N_j})\,,
\label{gamma}
\end{equation}
In Eq. (\ref{gamma}) each subfactor $\CS_{N_k}$ and ${\mathbb Z}_k$ permutes the $N_k$ cycles $(k)$ and acts within one cycle $(k)$
correspondingly. 
The total orbifold Hilbert space ${\cH}(\CS^N X)$ takes the form \cite{Dijkgraaf1}:
\begin{equation}
{\cH}(\CS^N X)=
\bigoplus_{[\gamma]} {\cH}_{\gamma}^{\Gamma_{\gamma}}
= \bigoplus_{[\gamma]}\bigotimes_{n \in {\mathbb Z}_+}
\CS^{N_n}{\cH}_{(n)}^{Z_{n}}\,,
\end{equation}
where each twisted sector ${\cH}_{\gamma}^{\Gamma_{\gamma}}$ has been decomposed into a product over the subfactors $(n)$ of $N_n$-fold symmetric tensor products, and $\CS^N{\cH}\equiv (\bigotimes_N{\cH})^{\CS_N}$.

Let $\chi({\bullet};q,y)$ be the generating function for every
(sub)Hilbert space of a supersymmetric sigma model. It has been
shown that the generating function coincides with the elliptic
genus \cite{Landweber}. For any vector bundle $E$ over a smooth
manifold we can define the formal sums
\begin{equation}
\wedge_q E = \bigoplus_{\atop{k \geq 0}}^d q^k\wedge^{k}E,\,\,\,\,\,
\CS_qE = \bigoplus_{\atop{k \geq 0}}^{\infty}q^k\CS^kE,\,\,\,\,\,
\chi_y(TX)=\bigoplus_{\atop{k=0}}^d(-y)^k\chi (\wedge^kTX)
\mbox{,}
\end{equation}
where $TX$ is the holomorphic tangent bundle of the K\"{a}hler
manifold $X$ of complex dimension $d$, $\wedge^k\,(\CS^k)$ and
denotes the k-th exterior (symmetric) product. The
Riemann-Roch-Hirzebruch theorem
\begin{equation}
\chi(X;q,y) \stackrel{{\rm def}}{=} \sum_j(-1)^j{\rm dim}\,H^j(X; E_{q,y})
=\int_{X}{\rm ch}\,(E_{q,y})Td (X)
\mbox{}
\label{RRH}
\end{equation}
gives an alternative definition of the elliptic genus
in terms of characteristic classes. The bundle
$E_{q,y}$ is
\begin{equation}
E_{q,y} = y^{-d/2}\bigotimes_{n\geq 1}\{\wedge_{-yq^{n-1}}TX
\bigotimes_{n\geq 1}^{\infty}\wedge_{-y^{-1}q^{n}}TX^{\prime}
\bigotimes_{n\geq 1}^{\infty}\CS_{q^n}TX
\bigotimes_{n\geq 1}^{\infty}\CS_{q^n}TX^{\prime}\}
\mbox{,}
\end{equation}
and therefore
\begin{eqnarray}
\chi(X;q,y) & = & \chi_y(TX) + q\bigoplus_{k=0}^d \left\{(-y)^{k+1}
\chi (\wedge^k TX\otimes TX)
+ q(-y)^{k-1}\chi (\wedge{}^k TX\otimes TX^{\prime})
\right.
\nonumber \\
& + & \left. q(-y)^{k}\chi (\wedge^kTX\otimes
(TX\otimes TX^{\prime}))\right\}
+{O}(q^{3/2})
\mbox{,}
\end{eqnarray}
where we have used the Riemann-Roch-Hirzebruch formula.
If $\chi({\cH}_{(n)}^{{\mathbb Z}_n};q,y)$ admits the extension
$\chi(\cH_{(n)}^{{\mathbb Z}_n};q,y)=\sum_{m\geq 0,\ell}\kappa(nm,\ell)q^my^{\ell}$,
the following result holds (see \cite{Kawai,Dijkgraaf2,Dijkgraaf1}):
\begin{eqnarray}
\sum_{N\geq 0}p^N\chi(\CS^N{\cH}_{(n)}^{{\mathbb Z}_n};q,y)
& = &
\prod_{m\geq 0,\ell}
\left(1-pq^my^{\ell}\right)^{-\kappa(nm,\ell)}\,,
\nonumber \\
\sum_{N\geq 0}p^N\chi(\CS^NX;q,y)
& = &
\!\!\!\!\prod_{n>0,m\geq 0,\ell}
\left(1-p^nq^my^{\ell}\right)^{-\kappa(nm,\ell)}\,,
\label{Id2}
\end{eqnarray}
$p:= e^{2\pi i\rho}$ and $y:= e^{2\pi iz}$,
$\rho$ and $\tau$ determine the complexified
K\"{a}hler form and complex structure modulus of ${T}^2$,
respectively, and $z$ parametrizes the $U(1)$ bundle on ${T}^2$.
\begin{remark}
The Narain duality
group $SO(3,2,{\mathbb Z})$ is isomorphic to the Siegel modular
group $Sp(4, {\mathbb Z})$ and it is convenient to combine
the parameters $\rho, \tau$ and a
Wilson line modulus $z$ into a $2\times 2$ matrix belonging to the
Siegel upper half-plane of genus two,
$\Xi = \left[
\begin{array}{ll}
\rho\,\,\,z &  \\
z\,\,\, \tau &
\end{array}
\!\!\!\!\!\!\!\right]$,
with ${\rm Im}\, \rho >0,\, {\rm Im}\, \tau >0$, ${\rm det}\,
{\rm Im}\, \Xi >0$.
The group $Sp(4, {\mathbb Z}) \cong SO(3,2,{\mathbb Z})$
acts on the $\Xi$ matrix by fractional linear transformations
$\Xi \rightarrow (A\Xi +B)(C\Xi + D)^{-1}$.
Note that for a Calabi-Yau space the $\chi_y-$genus
is a weak Jacobi form of zero weight and index
$d/2$ {\rm \cite{Hirzebruch1}} and it transforms as
$\chi_y (TX)=(-1)^{r-d}y^r\chi_{y^{-1}} (TX)$. This relation
can also be derived from the Serre duality 
\begin{equation}
H^j(X;{\wedge}^sTX)\cong
H^{d-j}(X;{\wedge}^{r-s}TX)\,.
\end{equation}
\end{remark}

In the case of the elliptic genus of a symmetric product space ${\mathfrak S}^NX$, if $y =1$ then the elliptic genus degenerates to the Euler number $\chi({\mathfrak S}^NX; q, 1)= {\bf e}({\mathfrak S}^NX)$
(or Witten index) \cite{Hirzebruch2,Vafa}.
For the symmetric product this gives the following identity, where
the character is almost a modular form of weight $-\chi(X)/2$\,:
\begin{equation}
\sum_{N\geq 0}q^N {\bf e}({\mathfrak S}^NX)=
\prod_{n \in {\mathbb Z}_+} \left(1-q^n\right)^{-\chi(X)} = 
[\cR(s= 1- it)]^{-\chi(X)}\,.
\label{character}
\end{equation}
In the case when $X$ is an algebraic surface, this formula also
computes the topological Euler characteristic of the Hilbert scheme $X^{[N]}$ of dimension zero subschemes of length $n$ 
\cite{Gottsche}.
A similar formula, associated to the (equivariant) orbifold Euler characteristic of the symmetric product, can be defined using the $\mathfrak{S}_N$-equivariant $K$-theory of $X^N$ by means of the following expression 
\footnote{
One can calculate the torsion free part of $K^\bullet_\Gamma(Y)$ (where $\Gamma$ acts on $Y$ and $\Gamma$ is a finite group) by localizing on the prime ideals of $R(\Gamma)$, the representation ring of $\Gamma$ (see for detail \cite{Segal})
$
\underline{K}^\bullet_\Gamma(Y)
\cong \bigoplus_{\{\gamma\}} \underline{K}^\bullet(Y^\gamma)^{\Gamma_\gamma},
$
where
$
\underline{K}_{\Gamma_N}(X^N)\equiv K_{\Gamma_N}(X^N)\otimes {\mathbb C}.
$
Here $\{\gamma\}$ runs over the conjugacy classes of elements in $\Gamma$, $Y^\gamma$ are the fixed point loci of $\gamma$ and $\Gamma_\gamma$ is the centralizer of $\gamma$ in $\Gamma$. The fixed point set $\{X^N\}^\gamma$ is isomorphic to $X^N = X^{\sum_nN_n}$, $\gamma \in {\mathfrak S}_N$, and $\Gamma_\gamma\cong \prod_n{\mathfrak S}_{N_n}\ltimes ({\mathbb Z}/n)^{N_n}$.
The cylic groups ${\mathbb Z} / n$ act trivially in $K^\bullet(X^{N})$ and therefore the following decomposition for the $\CS_N$-equivariant $K$-theory holds \cite{Lupercio}
\begin{equation}
\underline{K}^\bullet({\mathfrak S}^NX) \cong \bigoplus_{\{\gamma\}} \underline{K}^\bullet(({\mathfrak S}^N)^\gamma))^{\Gamma_\gamma} \cong \bigoplus_{\sum nN_n= N} \bigotimes_n \underline{K}^\bullet({\mathfrak S}^{N_n})^{\mathfrak{S}_{N_n}}\,.
\end{equation}
}
$
{\bf e}({\mathfrak S}^NX) := {\rm rank}\ K^0({\mathfrak S}^NX) - {\rm rank}\  K^1({\mathfrak S}^NX)\,.
$
\\

\noindent
{\bf Digression: Equivariant $K$-theory, wreath products.}
\label{Wreath}
We study here a direct sum of the equivariant $K$-groups $\cF_{\Gamma}(X):= \oplus_{N\geq 0} \underline{K}_{\Gamma_N}(X^N)$ associated with a topological $\Gamma$-space \cite{Wang}.  
$\Gamma$ is a finite group and the wreath (semi-direct) product 
$\Gamma_N\rtimes {\mathfrak S}_N$ acts naturally on the $N$th Cartesian product $X^N$. 
As an example of such $K$-group we shall analyze the group $K^{-}_{{\widetilde H}\Gamma_N}(X^N)$ which has been introduced in \cite{Wang}. The semi-direct product $\Gamma_N$ can be extended to the action of a larger finite supergroup 
${\widetilde H}\Gamma_N$, which is a double cover of the semi-direct product $(\Gamma\times {\mathbb Z}_2)^N\rtimes 
{\mathfrak S}_N$. The category of ${\widetilde H}\Gamma_N$-equivariant spin vector superbundles over $X^N$ is the category of $\Gamma_N$-equivariant vector bundles $E$ over $X^N$ such that $E$ carries a supermodule structure with respect to the complex Clifford algebra of rank $N$
\footnote{
A fundamental example of ${\widetilde H}\Gamma_N$-vector superbundles over $X^N$ ($X$ is compact) is the following:
given a $\Gamma$-vector bundle $V$ over $X$, consider the vector superbundle $V\oplus V$ over $X$ with the natural ${\mathbb Z}_2$-grading. One can endow the $N$th outer tensor product $(V\oplus V)^{\boxtimes N}$ with a natural ${\widetilde H}\Gamma_N$-equivariant vector superbundle structure over $X^N$.
}.
It has been shown \cite{Wang} that the following statements hold:
\begin{enumerate}[{-}]
\item{}
The direct sum $
\cF^{-}_{\Gamma}(X):= \oplus_{N=0}^{\infty} \underline{K}_{{\widetilde H}\Gamma_N}(X^N)
$ 
carries naturally a Hopf algebra structure. 
\item{}
It is isomorphic to the Fock space of a twisted Heisenberg superalgebra (in this section {\it super} means 
${\mathbb Z}_2$-graded) associated with 
$
K^{-}_{{\widetilde H}\Gamma_N}(X)\cong K_{\Gamma}(X).
$
\item{}
If $X$ is a point, the $K$-group 
$
\underline{K}_{{\widetilde H}\Gamma_N}(X^N)
$
becomes the Grothendieck group of spin supermudules of 
${\widetilde H}\Gamma_N$. 
\end{enumerate}
Such a twisted Heisenberg algebra has played an important role in the theory of affine Kac-Moody algebras \cite{Frenkel}. The structure of the space $\cF_{\Gamma}^{-}(X)$ under consideration can be modeled on the ring $\Omega_{\mathbb C}$ of symmetric functions with a linear basis given by the so-called Schur $Q$-functions (or equivalently on the direct sum of the spin representation ring of ${\widetilde H}\Gamma_N$ for all $N$). The graded dimension of the ring $\Omega_{\mathbb C}$ is given by the denominator $\prod_{n=0}(1-q^{2n-1})^{-1}$. 
We have mentioned in Sect. \ref{HH} that on the basis of
G\"{o}ttsche's formula \cite{Gottsche} it has been conjectured \cite{Vafa1} that the direct sum $\cH(S)$ of homology groups for Hilbert scheme $S^{[N]}$ of $N$ points on a (quasi-)projective surface $S$ should carry the structure of the Fock space of a Heisenberg algebra, which was realized subsequently in a geometric way \cite{Nakajima,Grojnowski}. Parallel algebraic structures such as Hopf algebra, vertex operators, and Heisenberg algebra as part of vertex algebra structures \cite{Borcherds1,Frenkel} have naturally showed up in $\cH(S)$ as well as in $\cF_{\Gamma}(X)$. If $S$ is a suitable resolution of singularities of an orbifold $X/\Gamma$, there appears close connections between $\cH(S)$ and $\cF_{\Gamma}(X)$ \cite{Wang}. In fact the special case of $\Gamma$ trivial is closely related to the analysis considered in \cite{Dijkgraaf}. It would be interesting to find some applications of results discussed above in string theory.

The orbifold Euler number ${\bf e}(X,\G)$ was introduced in
\cite{Dixon} in the study of orbifold string theory and it 
has been interpreted as the
Euler number of the equivariant $K$-group $K_{\Gamma}(X)$
\cite{Atiyah}. Define the Euler number of the generalized
symmetric product to be the difference
$
{\bf e}(X^N,{\widetilde H}\Gamma_N) := \dim K^{-,0}_{{\widetilde H}\Gamma_n}(X^N) - \dim
K^{-,1}_{{\widetilde H}\Gamma_N}(X^N)\,,
$
the series $\sum_{N=0}^{\infty} q^N {\bf e}(X^N,{\widetilde H}\Gamma_N)$ can be written in terms of spectral functions:   
\begin{eqnarray}
\sum_{N=0}^{\infty} q^N {\bf e}(X^N,{\widetilde H}\Gamma_N)  
& = &
\prod_{ n=1}^{\infty} (1-q^{2n-1})^{ -{\bf e}(X,\G)}
=
\left[q^{-\frac{25}{24}}(q-1)f_3(q)
\right]^{{\bf e} (X, \Gamma)}
\nonumber \\
& = &
\cR(s= 1/2 - (1/2)it)^{-{\bf e}(X, \Gamma)}
\,.
\end{eqnarray}

One can give an explicit description of ${\cF}^{-}_{\Gamma}(X)$ 
as a graded algebra.
Indeed, the following statement holds \cite{Wang}:             %
As a $(\Z_+ \times \Z_2)$-graded algebra, ${\cF}^-_{\G}(X, q)$
is isomorphic to the supersymmetric algebra
$ {\mathfrak S} \left( \bigoplus_{ N=1}^{\infty} q^{2N-1} \underline{K}_\G(X)\right)$. In particular, 
\begin{eqnarray}
\dim_q {\cF}^{-}_{\Gamma}(X) & = &
\prod_{ n=1}^{\infty}
\frac{(1 + q^{2n-1})^{ \dim K^1_\Gamma (X)} }
{(1 - q^{2n-1})^{ \dim K^0_\Gamma (X)}} 
\nonumber \\
& = &  
\frac{\left[\cR(\sigma=1/2 - (1/2)it + (1/2)i\eta(\tau))
\right]^{{\rm dim} K^1_\Gamma (X)}}
{\left[\cR(s=1/2 -(1/2)it)\right]^{{\rm dim} K^0_\G (X)}},
\end{eqnarray}
where the supersymmetric algebra is equal to the tensor product of
the symmetric algebra  
$ {\mathfrak S} \left( \bigoplus_{ N=1}^{\infty} q^{2N-1} \underline{K}^0_\Gamma (X)\right)$ and the exterior algebra 
$\Lambda \left( \bigoplus_{ N=1}^{\infty} q^{2N-1}
\underline{K}^1_\G(X) \right)$.
In the case when $X_{pt}$ is a point we have 
\begin{equation}
\sum_{N \geq 0} q^N \dim \cF^{-}_{\Gamma}(X_{pt})
= \prod_{ n=1}^{\infty} (1 -q^{2n-1})^{ -|\Gamma_*|} = 
\left[\cR(s=1/2 -(1/2)it)\right]^{ -|\Gamma_*|}\,.
\label{gamma00}
\end{equation}
In (\ref{gamma00}) $\Gamma$ is a finite group with $r+1$ conjugacy classes; $\Gamma^* := \{\gamma_j\}_{j=0}^r$ is the set of complex irreducible characters, where $\gamma_0$ denotes the trivial character. By $\Gamma_*$ we denote the set of conjugacy classes.

\subsection{Orbifold cohomology and the loop orbispace}
\label{Orbifold} 
For an orbifold $[Y/\Gamma]$ (viewed as a topological groupoid \cite{Moerdijk}) its {\it orbifold cohomology} is defined as the cohomology of the inertia orbifold $I[Y/\Gamma]$ \cite{Lupercio}, i.e. 
\begin{equation}
H_{orb}^\bullet([Y/\Gamma]) := H^\bullet(I[Y/\Gamma])\,.
\end{equation}
A description of the inertia orbifold of $[Y/\Gamma]$ can be given as follows:
$I[M/\Gamma] \cong \bigsqcup_{(\gamma)} [Y^\gamma/\Gamma_\gamma]$,
where $\{\gamma\}$ runs over the conjugacy classes, $Y^\gamma$ is the fixed point loci and $\Gamma_\gamma$ is the centralizer.
Then, 
\begin{equation}
H_{orb}^\bullet([Y/\Gamma]; \real) \cong \bigoplus_{\{\gamma\}} H^\bullet(Y^\gamma;\real)^{\Gamma_\gamma},\,\,\,\,\,\,\,\,\,
\underline{K}_\Gamma^\bullet(Y)\cong H_{orb}^\bullet([Y/\Gamma];\complex)\,.
\label{K}
\end{equation}
The last equation in (\ref{K}) is a consequence of 
the Chern character isomorphism.
One can define the Poincar\'e orbifold polynomial 
\begin{equation}
P_{orb}([Y/\Gamma],y):= \sum y^j 
{\rm rank}\,H_{orb}^j([Y/\Gamma]; {\mathbb R}) \equiv \sum y^j b_j^{orb}([Y/\Gamma])\,,
\end{equation} 
where $b^{orb}_j$ is the $j$-th orbifold Betti number.
For the symmetric product, viewed as an orbifold groupoid $[\CS^NX]$, it follows that
\begin{equation}
H_{orb}^\bullet([\CS^NX];\real) \cong \bigoplus_{\sum nN_n= N} \bigotimes_n H^\bullet(X^{N_n};\real)^{\mathfrak{S}_{N_n}}\,. 
\label{Horb}
\end{equation}
Then calculating the orbifold Poincar\'e polynomial yields
\begin{eqnarray} 
\label{L1}
\sum_{N=0}^\infty q^N  P_{orb}([\CS^NX],y)  
& = & 
\sum_{N=0}^\infty q^N \left(\sum_{\sum nN_n= N } \prod_n P(X^{N_n}/\mathfrak{S}_{N_n},y) \right) 
\\
& = & 
\sum_{N=0}^\infty  \left(\sum_{\sum nN_n=N } \prod_n          
(q^n)^{N_n} P(X^{N_n}/\mathfrak{S}_{N_n},y) \right) 
\\
& = & 
\prod_{n\in {\mathbb Z}_+} \left( \sum_{N=0}^\infty q^{nN} P(\CS^NX,y) \right)
\\
&=& 
\prod_j\prod_{n \in {\mathbb Z}_+} \frac{(1 + q^n y^{2j+1})^{b_{2j+1}(X)}}
{(1 - q^n y^{2j})^{b_{2j}(X)}}\,.
\end{eqnarray}
Finally when $y=\pm 1$, we have the formulae 
\begin{eqnarray}
\!\!\!\!\!\!\!\!\!\!\!\!\!\!\!\!\!\!\!\!\!\!
\sum_{N=0}^\infty q^N P_{orb}([\CS^NX],\pm 1) 
\!\! & = & \!\! 
\prod_j\prod_{n \in {\mathbb Z}_+} \left[\frac{(1 + q^n )^{b_{2j+1}(X)}}
{(1 - q^n)^{b_{2j}(X)}}\right]^{\pm 1}
\nonumber \\
\!\! & = & \!\!
\prod_j
\left[\frac{\cR(\sigma = 1-it + i\eta(\tau))^{b_{2j}(X)}}
{\!\!\!\!\cR(s=1-it)^{b_{2j+1}(X)}}\right]^{\pm 1}\,.
\label{generating1}
\end{eqnarray}
Note that for these formulae to be valid one only needs that the cohomology of $X$ is finitely generated at each $n$ \cite{Lupercio}.

\noindent
{\bf Loop orbifold.}
The loop orbifold $L[Y/\Gamma]$ associated with an orbifold $[Y/\Gamma]$ has been defined in \cite{Lupercio1,Lupercio2,Lupercio3}. 
For the case of a global quotient it has a simple description: $L[Y/\Gamma] = [\cP_\Gamma Y/\Gamma]$
\footnote{
The loop orbifold has another presentation (Morita equivalent) given by
$L[Y/G] \cong \bigsqcup_{(\gamma)} [\cP_\gamma Y/\Gamma_\gamma]$,\, where
$\Gamma_\gamma$ acts on $\cP_\gamma Y$ in the natural way.
},
\begin{equation}
\cP_\Gamma Y = \bigsqcup_{\gamma \in \Gamma} \cP_\gamma Y \times \{\gamma\}\,\,\,\,\,\,\, {\rm with}\,\,\,\,\, 
\cP_\gamma Y = \{ f \colon [0,1] \to Y | f(0)\gamma = f(1) \}
\end{equation} 
and the $\Gamma$ action is given by
\begin{eqnarray}
\Gamma \times \bigsqcup_{\gamma\in \Gamma} \cP_\gamma Y \times\{\gamma\} 
& \longrightarrow &  
\bigsqcup_{\gamma\in \Gamma} \cP_\gamma Y \times\{\gamma\},
\nonumber \\
(h, (f,\gamma))  
& \longmapsto &  
(f \cdot h, h^{-1}\gamma h)\,
(f\cdot h(t) := f(t)h).
\end{eqnarray}
A lemma, which has been proved in \cite{Lupercio2}, asserts the
following isomorphism:
\begin{equation} 
[\cP_\gamma X^N /\Gamma_\gamma] \, \cong \,
\bigotimes_n \,[({\mathfrak L} X)^{N_n}/{\mathfrak S}_{N_n}
\ltimes({\mathbb Z}/n )^{N_n}]\,,
\end{equation} 
where the action of ${\mathbb Z}/ n$ is given by rotation by the angles $2\pi k/ n$ on ${\mathfrak L} X$, the free loop space of $X$. Because of the action of ${\mathbb Z}/n$ in ${\mathfrak L}X$ factors through the rotation action of the cycle $S^1$ in 
${\mathfrak L}X$, it follows (see for detail \cite{Lupercio})
\begin{equation}
H^\bullet(L[\CS^NX]; {\mathbb Z})
\cong \bigoplus_{\{\gamma\}} H^\bullet(\cP_\gamma X^N ; {\mathbb Z})^{\Gamma_\gamma} \cong \bigoplus_{\sum nN_n=N}\bigotimes_n H^\bullet(({\mathfrak L} X)^{N_n} ; {\mathbb Z})^{{\mathfrak S}_{N_n}}\,.
\label{L2}
\end{equation}
Let $X$ be such that $H^j( {\mathfrak L} X;\real)$ is finitely generated. Then using the previous results, Eqs. (\ref{Horb}), (\ref{L1}), (\ref{L2}),
one can get
\begin{eqnarray}
\sum_{N=0}^\infty q^N P({L}[\CS^NX],y) & = & \prod_{j} 
\prod_{n\in {\mathbb Z}_+}\frac{[1 + q^n y^{2j+1}]^{b_{(j\,\,{\rm odd})}   ({\mathfrak L} X)}}{[1 - q^n y^{2j}]^{b_{(j\,\,{\rm even})}({\mathfrak L} X)}}
\nonumber \\
& = &
\prod_j \frac{[\cR (\sigma = (2j+1)(1-it){\rm log}\,y + i\eta(\tau))]^{b_{(j\,\,{\rm odd})}({\mathfrak L} X)}}
{[\cR (s = 2j(1-it){\rm log}\,y )]^{b_{(j\,\,{\rm even})}({\mathfrak L} X)}}
\label{loop1}
\end{eqnarray} 
where $b_{j}({\mathfrak L} X)$ is the $j$-th Betti number of 
${\mathfrak L} X$. When $y= \pm 1$ Eq. (\ref{loop1}) leads to formula (\ref{generating1}). Via the Chern character map we get
\begin{equation}
\underline{K}^\bullet_{\CS_N}(({\mathfrak L} X)^N) \cong H^\bullet({L} [\CS^NX];\complex)\,.
\end{equation}
The fact that the cohomologies of $I[{\mathfrak L} \CS^NX] $ and  
${L} [\CS^NX]$ agree is a feature of the symmetric product. In general, for any orbifold $[Y/G]$, the cohomologies of $I[{\mathfrak L} Y /\Gamma]$ and ${L} [Y/\Gamma]$ do not necessarily agree.

\subsection{Symmetric products of $K3$} 
\label{K3}
Let a five-brane be wrapped around a $K3$ compactification manifold so that its world-volume has the space-time topology of 
$K3\times S^1$. The space $K3$ has only non-trivial homology cycles of even dimension, so to a given $D$-brane configuration we can associate an element of the vector space 
$H_{\bullet}(K3; {\mathbb Z})$ in the integral homology of $K3$.
Such a vector can be identified with the charge vector via the isomorphism \cite{Dijkgraaf1}: $H_{\bullet}(K3; {\mathbb Z})\cong
\Gamma^{20, 4}$, in which the norm on 24-lattice $\Gamma^{20, 4}$
is defined with the intersection form. Note that a pair of three-branes can intersect along one-dimensional string wrapping one (or more times)
around the $S^1$ direction, while the one-brane intersects with the five-brane along a string around the $S^1$. Thus these strings formed by the $D$-brane intersections encode the internal degrees of freedom of the five-brane
\footnote{
In type II string compactifications on manifolds of the form $X\times S^1$ it is possible to consider the configuration of a $D$-string wound $N$ times around the $S^1$, bound to a $({\rm dim} X +1)$-brane. In the case when $X= K3$ this situation has been considered in \cite{Strominger1} for $D$-brane computation of the five-dimensional black hole entropy. The quantum mechanical degrees of friedom of this $D$-brane configuration are naturally encoded in terms of a two-dimensional sigma model on the $N$-fold symmetric tensor product of $X$, that describes the transversal fluctuations of the $D$-string.
}.
It has been shown that such multiple $D$-brane description leads to a two-dimensional sigma model with target space the sum of symmetric products of $K3$ space \cite{Strominger1}. 

The counting of the BPS states is naturally related to the elliptic genus of this sigma model while the degeneracies are given in terms of the denominator of a generalized super Kac-Moody algebra 
\footnote{
A denominator formula can be written as follows (in \cite{Borcherds92} the product formulas of type (\ref{Id2}) has been interpreted in terms of the so-called denominator formula for the monster Lie algebra) \cite{Borcherds,Harvey}:
$$
\sum_{\sigma\in {W}}\left({\rm sgn}(\sigma)\right)e^{\sigma(v)}
=e^{v}\prod_{n\in {\mathbb Z}_+}\left(1-e^{n}\right)^{{\rm mult}(n)},
$$
where $v$ is the Weyl vector, the sum on the left hand side is over all elements of the Weyl group ${W}$, the product on the right
hand side runs over all positive roots (one has the usual notation of root spaces, positive roots, simple roots and Weyl group, associated with Kac-Moody algebra) and each term is weighted by the root multiplicity
${\rm mult}(n)$.
}.
Remember that in the case of an affine Lie algebra the generalized Weyl denominator identity turns out to be equivalent to the Macdonald identities; historically this was the first application of the representation theory of Kac-Moody algebras \cite{Kac}. 
The generating function becomes \cite{Vafa}:
\begin{equation}
\sum_N q^{N}{\rm dim}\, H^\bullet({\mathfrak S}^NK3=(K3)^N/{\mathfrak S}_N) 
=\prod_{n\in {\mathbb Z}_+}(1-q^n)^{-24}
= [\cR(s=1-it)]^{-24}\,.
\label{GF1}
\end{equation}
In the case of a $K3$ or an abelian surface, the orbifold elliptic genus of the symmetric product also coincides with elliptic genus of the Hilbert scheme.

\section{Conclusions}
\label{Conclusions}

In this paper we have discussed how elliptic modular forms and spectral functions of $AdS_3$-asymptotic geometry are intertwined with quantum generating functions of gravity (and black holes), and with elliptic genera of sigma models associated to generalized symmetric products.
It is of course of utmost importance to find generalizations of the example considered here. Perhaps one could make analogues of $N$-fold products also for other type of supergravity solutions (see for example \cite{Bytsenko00}). Indeed, we recall that new classes of eleven-dimensional supergravity warped product of $AdS_3$ with an eight-dimensional manifold $X_8$, which are dual to conformal field theories with $N=(0,2)$ supersymmetry (since of the $AdS/CFT$ correspondence), have been found in \cite{Waldram2}. These new solutions are all $S^2$ bundles over six-dimensional base spaces. A two-sphere bundle can be obtained from the canonical line-bundle over a six-dimensional base space $\cB_6$. Physically this class of solutions might be useful in
obtaining type IIB and eleven-dimensional bubble solutions \cite{Kim1,Kim2} and it might be derived from the multi-charged superstars \cite{Waldram2}. Spaces $\cB_6$ are products of
K\"ahler-Einstein $N$-spaces ($KE_N$) with various possibilities for the signs of the curvature.

To obtain new $N$-fold solutions one can consider, for example, a general class of K\"ahler-Einstein spaces of the form $X_{2N} = KE_2^{(1)}\times \cdots \times KE_2^{(N)}$, where $KE_2$ is a two-dimensional space of negative curvature (the reader can find in  \cite{Bonora,Bytsenko} some homological and $K$-theory methods applied to hyperbolic cycles). 
This class includes the product of $2N$-fold two-space forms, which are associated to discrete subgroups of $SL(2, {\mathbb R})^N$ with compact quotient 
${\bf H}^N/\Gamma \equiv (\underbrace{
{H}^2\times \ldots \times{H}^2}_N)/\Gamma$ \cite{BytsenkoNew}. In this context one could consider the following central topics:
{\bf (i)}\, A discrete subgroup $\Gamma\subset SL(2, {\mathbb R})^N$ with compact quotient ${\bf H}^N/\Gamma$ and the Hilbert modular group. (The corresponding singular cohomology groups $H^\bullet (\Gamma; {\mathbb C})$ have been determined in \cite{Matsushima}).
{\bf (ii)}\, The Eilenberg-MacLane cohomology groups
$ H^{\bullet}(\Gamma_{k}; {\mathbb C})$, where $\Gamma_{k}$ acts trivially on $\mathbb C$ and the totally real number field 
${k}\supset {\mathbb Q}$. The corresponding spaces
are the {\it Hilbert modular varieties}:
${\bf H}^N/\Gamma_{\bf k}$.  
{\bf (iii)}\, The mixed Hodge structure in the sense of Deligne \cite{Deligne}.
The Hilbert modular group, $\Gamma_{k}=SO(2, {\bf o})$, and the corresponding spaces (Hilbert modular varieties) and functions ({\it Hilbert modular forms}) have been actively studied in mathematics (for reference see \cite{Freitag}).
$\Gamma_{k}$ is the group of all $2\times 2$ matrices of determinant 1 with coefficients in the ring $\bf o$ of integers of a totally real number field. The Eilenberg-MacLane cohomology groups
$H^\bullet(\Gamma_{k}; {\mathbb C})$ are isomorphic to the singular cohomology group of the Hilbert modular variety
$
H^\bullet (\Gamma_{k}; {\mathbb C}) =
H^\bullet ({\bf H}^N/\Gamma_{k};
{\mathbb C}),
$
where, as before, ${\bf H}^N$ denotes the product of $N$ upper half-planes $H^2$ equipped with the natural action of 
$\Gamma_{k}$. The Hilbert modular variety carries a natural structure as a {\it quasiprojective variety} and its cohomology groups inherit a {\it Hodge structure}. In fact the Hilbert modular group is a simplified example of the cohomology theory of arithmetic groups
and it is the only special case in which the cohomology can be determined explicitly.

Having advocated in this paper the basic role of Lie algebra homologies and cohomologies, we are naturally led to other problems, related to the quantization of nonlinear sigma models and gravity. For instance, one might ask whether 
the quantum behaviour of these models can be algebraically intepreted by means of infinitesimal deformations of the corresponding Lie algebra $\mathfrak g$. No doubt this analysis requires a new degree of mathematical sophistication. 
Perhaps all the concepts of what should be the ``deformation theory of everything" might be tested in the case of associative algebras, which are algebras over operads \cite{Kontsevich00}. { In many examples dealing with algebras over operads, arguments of the universality of associative algebras are called forth. This may suggest that a connection between the deformation theory and algebras over operads.}
The space $H^2({\mathfrak g}; {\mathfrak g})$, i.e. the cohomology of the algebra $\mathfrak g$ with coefficients in the adjoint representation, may be interpreted as the set of classes of infinitisimal deformations of the algebra $\mathfrak g$. In this connection
infinitisimal deformations are cocycles from the cochain complex
$C^2({\mathfrak g}; {\mathfrak g})$, while equivalent infinitisimal deformations are cohomologous cocycles. 
Suppose $[\vartheta]$ is a class of infinitisimal deformations of $\mathfrak g$. 
For the existence of the deformation it is necessary that all the Massey powers (i.e. at least the elements of the space 
$H^3({\mathfrak g}; {\mathfrak g})$) of the class $[\vartheta]$ vanish \cite{Fuks}.
\footnote{
Recall the higher multiplicative structures due to Massey. The simplest multiplication is the operator which assignes to every triple
$
\alpha\in H^m(\mathfrak{g}),\, \beta\in H^n(\mathfrak{g}),\,
\gamma\in H^k(\mathfrak{g})
$
such that
$
\alpha\beta =0,\, \beta\gamma = 0 
$
an element $\langle \alpha, \beta, \gamma\rangle$ of the form
$
\langle \alpha, \beta, \gamma\rangle :\, 
H^{m+n+k-1}(\mathfrak{g})/[\alpha H^{n+k-1}(\mathfrak{g})+ H^{m+n-1}(\mathfrak{g})\gamma].
$
The next Massey operation is defined on quadruples
$
\alpha\in H^m(\mathfrak{g}),\, \beta\in H^n(\mathfrak{g}),\,
\gamma\in H^k(\mathfrak{g}),\, \delta\in H^\ell(\mathfrak{g})
$
such that
$\alpha\beta =0,\, \beta\gamma = 0 ,\, \gamma\delta = 0
$
and
$ 
\langle \alpha, \beta, \gamma\rangle =0,\,
\langle \beta, \gamma, \delta \rangle =0\,.
$
This operation assumes values in the quotient space
$
\langle \alpha, \beta, \gamma, \delta \rangle : \,
H^{m+n+k+\ell-2}(\mathfrak{g})/[H^{m+n-1}(\mathfrak{g})
H^{k+\ell-1}(\mathfrak{g}) + \alpha H^{n+k+\ell-2}(\mathfrak{g})\gamma
+H^{m+n+k-2}(\mathfrak{g})\delta]\,.
$
Note that multiplicative structure arises in the space $H^\bullet({\mathfrak g}; {\mathfrak g})$, and with some suitable relations in cochains $C^\bullet({\mathfrak g}; {\mathfrak g})$ one can get the multiplication in cohomology 
$
H^m({\mathfrak g}; {\mathfrak g})\otimes
H^n({\mathfrak g}; {\mathfrak g})\rightarrow
H^{m+n-1}({\mathfrak g}; {\mathfrak g}).
$
This multiplication (it reminds the topologist of the Whitehead product in homotopy groups) supplies the space $H^\bullet({\mathfrak g}; {\mathfrak g})$ with a Lie superalgebra structure, and it generates the sequence of Massey products.
We refer the reader to book \cite{Fuks} for futher details.
}
Ordinary multiplication, as well as Massey multiplication, can be naturally related to the class of the infinitisimal deformations of Lie algebras and therefore to deformations of generating functions of the quantum theory. We hope we will be able to discuss this problem in forthcoming papers.

\subsection*{Acknowledgements}
The authors would like to thank Professors M. E. X. Guimar\~aes and J. A. Helay\"{e}l-Neto for useful discussions. L.B. would like to thank ICTP and the University of Londrina for financial support.
A. A. Bytsenko would like to acknowledge the Conselho Nacional de Desenvolvimento Cient\'ifico e Tecnol\'ogico (CNPq) for support. 
Research of A. A. Bytsenko was performed in part while on leave at the High Energy Physics Group of ICTP and the High Energy Sector of SISSA, Trieste, Italy.


\begin{thebibliography}{0}




\bibitem{Brown}
J. D. Brown and M. Henneaux, {\it Central Charges in the Canonical Realization of Asymptotic Symmetries: An Example from Three-Dimensional Gravity}, Commun. Math. Phys. {\bf 104} (1986) 207.

\bibitem{Giombi}
S. Giombi, A. Maloney and X. Yin, {\it One-Loop Partition Functions of 3D Gravity}, JHEP {\bf 0808} (2008) 007;  [arXiv:hep-th/0804.1773].

\bibitem{Fuks}
D. B. Fuks, {\it Cohomology of Infinite-Dimensional Lie Algebras},
Contemporary Soviet Mathematics, Consultans Bureau, New York,
1986.


\bibitem{Ochanine}
S. Ochanine, {\it Sur les genres multiplicatifs definis par des integrales elliptiques}, Topology {\bf 26} (1987) 143. 

\bibitem{Witten00}
E. Witten, {\it Elliptic Genera and Quantum Field Theory},
Commun. Math. Phys. {\bf 109} (1987) 525.

\bibitem{Landweber1}
P. S. Landweber, {\it Elliptic cohomology and modular forms}, in P. S. Landweber, editor, Elliptic Curves and Modular Forms in Algebraic Topology (Proceedings, Princeton 1986), Lecture Notes in Math. {\bf 1326} 55–68, Springer, 1988.

\bibitem{Landweber2}
P. S. Landweber, D. Ravenel, and R. Stong, {\it Periodic cohomology theories defined by elliptic curves}, In The \u{C}ech Centennial
(Boston, MA, 1993), Contemp. Math. {\bf 181}, Amer. Math. Soc., Providence, RI, 1995.

\bibitem{Wang}
W. Wang, {\it Equivariant $K$-theory, generalized symmetric products, and twisted Heisenberg algebra}, 
Commun. Math. Phys. {\bf 234} (2003) 101; [arXiv:math.QA/0104168v2].

\bibitem{Strominger1}
A. Strominger and C. Vafa, {\it Microscopic Origin of the Bekenstein-Hawking Entropy}, Phys.  Lett. 
B {\bf 379} (1996) 99; [arXiv:hep-th/9601029].

\bibitem{Perry}
P. Perry and F. Williams, {\it Selberg zeta function and trace formula for the BTZ black hole}, Internat. J. of Pure and Applied Math. {\bf 9} (2003) 1.

\bibitem{Bytsenko4}
A. A. Bytsenko, M. E. X. Guim\~araes and F. L. Williams, 
{\it Remarks on the Spectrum and Truncated Heat Kernel of the BTZ Black Hole}, Lett. Math. Phys. {\bf 79} (2007) 203; 
[arXiv:hep-th/0609102].

\bibitem{Bytsenko1}
A. A. Bytsenko and M. E. X. Guim\~araes, {\it Expository Remarks on Three-Dimensional Gravity and Hyperbolic Invariants}, 
Class. Quantum Grav. {\bf 25} (2008) 228001;
[arXiv:hep-th/0809.5179]. 

\bibitem{Kac}
V. G. Kac, {\it Infinite dimensional Lie algebras}, Cambridge University Press, 1990.

\bibitem{Fried}
D. Fried, {\it Analytic torsion and closed geodesics on hyperbolic manifolds}, Invent. Math. {\bf 84} (1986) 523.

\bibitem{Floyd}
F. L. Williams, {\it Vanishing Theorems for Type $(0, q)$ Cohomology of Locally Symmetric Spaces}, Osaka J. Math. {\bf 18}
(1981) 147.

\bibitem{Abdalla}
M. C. B. Abdalla, A. A. Bytsenko and M. E. X. Guimar\~{a}es,
{\it Complex Topological Invariants of Three-Hyperbolic Manifolds},
J. Dinamical Systems and Geometric Theories {\bf 4} (2006) 1.

\bibitem{Deitmar}
A. Deitmar, {\it The Selberg Trace Formula and the Ruelle Zeta Function for Compact Hyperbolics}, Abh. Math. Se. Univ. Hamburg {\bf 59} (1989) 101.

\bibitem{Perchik}
J. Perchik, {\it Cohomology of Hamiltonian and related formal vector fields Lie algebras}, Topology {\bf 15} (1976) 395.

\bibitem{Andrews}
G. E. Andrews, {\it The Theory of Partitions}, In Encyclopedia of Mathematics and its Applications, Addison-Wesley Publishing Company, 1976.

\bibitem{ElizaldeBook}
E. Elizalde, S. D. Odintsov, A. Romeo, A. A. Bytsenko and S. Zerbini,  {\it Zeta Regularization Techniques with Applications}, World
Sci., Singapore, 1994.

\bibitem{BytsenkoPhysRep}
A. A. Bytsenko, G. Cognola, L. Vanzo and S. Zerbini, {\it Quantum Fields and Extended Objects in Space-Times with Constant Curvature Spatial Section}, Phys. Rept. {\bf 266} (1996) 1-126.

\bibitem{Maloney}
A. Maloney and E. Witten, {\it Quantum Gravity Partition
Function In Three Dimensions}, JHEP {\bf 1002} (2010) 029; [arXiv:hep-th/0712.0155].

\bibitem{Bytsenko2}
A. A. Bytsenko and M. E. X. Guim\~araes, {\it Truncated Heat Kernel and One-Loop Determinants for the BTZ Geometry}, Eur. Phys. J. 
C {\bf 58} (2008) 511; [arXiv:hep-th/0809.1416]. 

\bibitem{Bytsenko3}
A. A. Bytsenko and M. E. X. Guim\~araes, {\it Partition Functions of Three-Dimensional Quantum Gravity and the Black Hole Entropy},
J. Phys. Conf. Ser. {\bf 161} (2009) 012023; [arXiv:hep-th/0807.2222].

\bibitem{Perry1}
P. Perry, {\it A Poisson summation formula and lower bounds for resonances in hyperbolic manifolds}, Int. Math. Res. Notes
{\bf 34} (2003) 1837.

\bibitem{Patterson1}
S. J. Patterson, {\it The Selberg zeta-function of a Kleinian group}, In Number Theory, Trace Formulas, and Discrete Groups: Symposium in Honor of Atle Selberg, Oslo, Norway, July 14-21, 1987, New York, Academic Press, 1989.

\bibitem{Patterson2}
S. J. Patterson and P. A. Perry, {\it The divisor of the Selberg zeta function for Kleinian groups, with an appendix by Charles Epstein}, Duke Math. J. {\bf 106} (2001) 321. 

\bibitem{Belavin1}
A. A. Belavin, A. M. Polyakov and A. B. Zamolodchikov, {\it Infinite conformal symmetry of critical fluctuations in two dimensions}, J. Stat. Phys. {\bf 34} (1984) 763.

\bibitem{Belavin2}
A. A. Belavin, A. M. Polyakov and A. B. Zamolodchikov, {\it Infinite conformal symmetry in two-dimensional quantum field theory}, Nucl. Phys. B
{\bf 241} (1984) 333.

\bibitem{Dijkgraaf3}
R. Dijkgraaf, {\it Discrete torsion and symmetric products}, 
arXiv:hep-th/9912101.
 
\bibitem{Kleban}
M.~Kleban, M.~Porrati and R.~Rabadan,
{\it Poincar\'{e} Recurrences and Topological Diversity},
JHEP {\bf 0410} (2004) 030; [arXiv:hep-th/0407192].

\bibitem{Manschot}
J.~Manschot, {\it $AdS_3$ Partition Functions Reconstructed}, JHEP {\bf 0710} (2007) 103; [arXiv:hep-th/0707.1159].

\bibitem{Iwaniec}
H. Iwaniec, {\it Spectral Methods of Authomorphic Forms}, (Graduate Studies in Mathematics, {\bf 53}), Amer. Math. Soc. 2002.

\bibitem{Macdonald}
I. G. Macdonald, {\it The Poincar\'{e} polynomial of a symmetric product}, Proc. Cambridge Philos. Soc. {\bf 58} (1962) 563.

\bibitem{Macdonald1}
I. G. Macdonald, {\it Affine root systems and Dedekind's function},
Invent. Math. {\bf 15} (1972) 91.

\bibitem{Serre}
J.-P. Serre, {\it Lie Algebras and Lie Groups}, Lecture Notes in
Mathematics, {\bf 1500}, Springer-Verlag, 1992.

\bibitem{Nakajima00}
N. Nakajima, {\it Lectures on Hilbert schemes of points on surface}, ANS Univ. Lectures Series {\bf 18}, 1999.

\bibitem{Gottsche}
L. G\"{o}ttsche, {\it The Betti numbers of the Hilbert Scheme of Points on a Smooth Projective Surface}, Math. Ann. {\bf 286} (1990) 193.

\bibitem{Kac2}
V. Kac and D. Peterson, {\it Infinite dimensional Lie algebras, theta functions and modular forms}, Advances in Math. {\bf 53}
(1984) 125.

\bibitem{Fogarty}
J. Fogarty, {\it Algebraic families on an algebraic surface}, Amer. J. Math. {\bf 90} (1968) 511.

\bibitem{Fujiki}
A. Fujiki, {\it On primitive symplectic compact K\"{a}hler v-manifolds of dimension four, in ''Classification of Algebraic and Analytic Manifolds``}, K. Ueno (ed.), Progress in Mathematics, Birkh\"{a}user {\bf 39} (1983) 71. 

\bibitem{Beauville}
A. Beauville, {\it Vari\'{e}t\'{e} k\"{a}hleriennes dont la premi\'{e}re classe de Chern est nulle}, J. of Differential Geom.
{\bf 18} (1983) 755.

\bibitem{Dijkgraaf1}
R. Dijkgraaf, G. Moore, E. Verlinde and H. Verlinde,
{\it Elliptic Genera of Symmetric Products and Second Quantized
Strings}, Commun. Math. Phys. {\bf 185} (1997) 197;
[arXiv:hep-th/9608096].

\bibitem{Vafa95}
C. Vafa, {\it Instantons on D-branes}, Nucl. Phys. B {\bf 463} (1996) 435; [arXiv:hep-th/9512078].

\bibitem{Bershadsky95}
M. Bershadsky, V. Sadov and C. Vafa, {\it D-Branes and Topological Field Theories}, Nucl. Phys. B {\bf 463} (1996) 420; 
[arXiv:hep-th/9511222].

\bibitem{Kontsevich}
M. Kontsevich, {\it Enumeration of rational curves via torus actions},
The moduli space of curves (Texel Island, 1994), Progr. Math. {\bf 129} 335-368, Birkh\"{a}user Boston, Boston, MA, 1995; MR: MR1363062 (97d:14077).

\bibitem{Moerdijk}
I. Moerdijk, {\it Orbifolds as groupoids: an introduction}, Orbifolds in mathematics and physics (Madison, WI, 2001), Contemp.
Math. {\bf 310}, Amer. Math. Soc., Providence, RI (2002) 205, MR 1 950 948. 

\bibitem{Dixon}
L. Dixon, J.A. Harvey, C. Vafa, and E. Witten, {\it Strings on
orbifolds}, Nucl. Phys. B {\bf 261} (1985) 678.

\bibitem{Landweber}
P. S. Landweber (Editor), {\it Elliptic Curves and Modular Forms in Algebraic Topology}, Proceedings of a Conference held at the Institute for Advanced Study, Princeton; Lecture Notes in Mathematics {\bf 1326}, Springer-Verlag, 1988.

\bibitem{Kawai}
T. Kawai, Y. Yamada and S.-K. Yang,
{\it Elliptic Genera and N=2 Superconformal Field Theory},
Nucl. Phys. B {\bf 414} (1994) 191; [arXiv:hep-th/9396996].

\bibitem{Dijkgraaf2}
R. Dijkgraaf, E. Verlinde and H. Verlinde, {\it Counting Dyons in N=4 String Theory}, Nucl. Phys. B {\bf 484} (1997) 543;
[arXiv:hep-th/9607026].

\bibitem{Hirzebruch1}
F. Hirzebruch, {\it Topological Methods in Algebrtaic Geometry},
Springer-Verlag (3rd Edition), 1978.

\bibitem{Hirzebruch2}
F. Hirzebruch and T. H\"{o}fer, {\it On the Euler Number of an
Orbifold},  Math. Ann. {\bf 286} (1990) 255.

\bibitem{Vafa}
C. Vafa, {\it Black Holes and Calabi-Yau Threefolds},
Adv. Theor. Math. Phys. {\bf 2} (1998) 207; [arXiv:hep-th/9711067].

\bibitem{Segal}
G. Segal, {\it Equivariant K-theory}, Inst. Hautes \'{E}tudes Sci. Publ. Math. {\bf 34} (1968) 129; MR MR0234452 (38 $\#$ 2769).

\bibitem{Lupercio}
E. Lupercio, B. Uribe, and M. A. Xicotencatl, {\it The Loop Orbifold of the Symmetric Product}, 
J. Pure and Applied Algebra, ISSN 0022-4049 {\bf 211}
(2007) 293; [arXiv:math.AT/0606573].

\bibitem{Frenkel}
I. Frenkel, J. Lepowsky and A. Meurman, {\it Vertex operator algebras and the Monster}, Academic Press, New York 1988.

\bibitem{Vafa1}
C. Vafa and E. Witten, {\it A strong coupling test of S-duality},
Nucl. Phys. {\bf 15} (1995) 189.

\bibitem{Nakajima}
H. Nakajima, {\it Lectures on Hilbert schemes of points on surfaces}, Univ. Lect. Ser. {\bf 18}, Amer. Math. Soc., 1999.

\bibitem{Grojnowski}
I. Grojnowski, {\it Instantons and affine algebras I: the Hilbert scheme and vertex operators}, Math. Res. Lett. {\bf 3} (1996) 275.

\bibitem{Borcherds1}
R. E. Borcherds, {\it Vertex algebras, Kac-Moody algebras, and the Monster}, Proc. Natl. Acad. Sci. USA {\bf 83} (1986) 3068.

\bibitem{Dijkgraaf}
J. de Boer, M. C. N. Cheng, R. Dijkgraaf, J. Manschot and E. Verlinde, {\it A Farey Tail for Attractor Black Holes},
JHEP {\bf 0611} (2006) 024; [arXiv:hep-th/0608059].

\bibitem{Atiyah} 
M. Atiyah and G. Segal, {\it On equivariant Euler characteristics}, J. Geom. Phys. {\bf 6} (1989) 671.

\bibitem{Lupercio1}
E. Lupercio and B. Uribe, {\it Loop groupoids, gerbes, and twisted sectors on orbifolds}, Orbifolds in mathematics and Physics (Madison, WI 2001), Contemp. Math. {\bf 310} (2002) 163, Amer. Math. Soc., Providence, RI MR 1 950 946. 

\bibitem{Lupercio2}
E. Lupercio, B. Uribe, and M. A. Xicotencatl, {\it Orbifold string topology}, Geom. Topol. {\bf 12} (2008) 2203;  [arXiv:math.AT/0512658].

\bibitem{Lupercio3}
E. Lupercio, B. Uribe, and  A. Xicotencatl, {\it Topological quantum field theories, strings, and orbifolds}, Contemp.
Math. {\bf 434} (2007) 73;  [arXiv:hep-th/0605255].

\bibitem{Borcherds92}
R. E. Borcherds, {\it Monstrous moonshine and monstrous Lie superalgebras}, Invent. Math. {\bf 109} (1992) 405.

\bibitem{Borcherds}
R. E. Borcherds, {\it Automorphic Forms on ${\rm O}\sb
{s+2,2}(R)$ and Infinite Products}, Invent. Math. {\bf 120}
(1995) 161.

\bibitem{Harvey}
J. A. Harvey and G. Moore, {\it Algebras, BPS States, and Strings},
Nucl. Phys. B {\bf 463} (1996) 315; [arXiv:hep-th/9510182].

\bibitem{Bytsenko00}
A. A. Bytsenko, M. E. X. Guimar\~aes and J. A. Helay\"{e}l-Neto,
{\it Hyperbolic Space Forms and Orbifold Compactification in M-Theory}, PoS WC2004 (2004) 017; [arXiv:hep-th/0502031]. 

\bibitem{Waldram2}
J. P. Gauntlett, N. Kim and D. Waldram,
{\it Supersymmetric $AdS_3, AdS_2$ and Bubble Solutions},
JHEP {\bf 0704} (2007) 005; [arXiv:hep-th/0612253].

\bibitem{Kim1}
N.~Kim, {\it AdS(3) solutions of IIB supergravity from D3-branes},
JHEP {\bf 0601} (2006) 094; [arXiv:hep-th/0511029].

\bibitem{Kim2}
N.~Kim and J.~D.~Park, {\it Comments on AdS(2) solutions of D = 11 supergravity}, JHEP {\bf 0609} (2006) 041;
[arXiv:hep-th/0607093].

\bibitem{Bonora}
L. Bonora and A. A. Bytsenko, {\it Fluxes, brane charges and Chern morphisms of hyperbolic geometry}, Class. Quantum Grav. {\bf 23} (2006) 3895; [arXiv:hep-th/0602162].

\bibitem{Bytsenko}
A. A. Bytsenko, {\it Homology and K-Theory Methods for Classes of Branes Wrapping Nontrivial Cycles}, J. Phys. A: Math. and Gen. {\bf 41} (2008) 045402; [arXiv:hep-th/0710.0305].

\bibitem{BytsenkoNew}
A. A. Bytsenko and E. Elizalde, {\it AdS Solutions in Gauge Supergravities and Global Anomaly for the Product of Complex Two-Cycles}, Eur. Phys. J. C {\bf 71} (2011) 1592.

\bibitem{Matsushima}
Y. Matsushima and G. Shimura, {\it On the cohomology groups attached to certain vector valued differential forms on the product of the upper half planes}, Ann. Math. {\bf 78}
(1963) 417.

\bibitem{Deligne}
P. Deligne, {\it Th\'eorie de Hodge. I, II}, Publ. Math., I.H.E.S.
{\bf 40} (1971) 5.

\bibitem{Freitag}
E. Freitag, {\it Hilbert Modular Forms}, Springer-Verlag, 1990.

\bibitem{Kontsevich00} 
M. Kontsevich and Y. Soibelman, {\it Deformations of algebras over operads and Deligne's conjecture}, arXiv:math/0001151v2.


\end{thebibliography}
\end{document}